\documentclass{ar2e}
\begin{document}
\input epsf.tex    
\input epsf.def   

\input psfig.sty

\def\ho{H\"oflich}
\def\Ha{H$\alpha$\,}
\def\Hb{H$\beta$\,}
\def\etal{et al.}
\def\kms{$\rm{km\,s^{-1}}$}
\def\arcsec{$^{\prime\prime}$}
\def\ergs{erg\,s$^{-1}$}
\def\Po{$P_{\rm o}\ $}
\def\degree{$^{\rm o}$}
\def\CaII7291{Ca {\sc II}] $\lambda\lambda$ 7291,7323\ }
\def\HeIA{He {\sc I} $\lambda$ 5876\ }
\def\HeIB{He {\sc I} $\lambda$ 6678\ }
\def\OI6300{[O {\sc I}] $\lambda\lambda$ 6300,6364\ }
\def\OIA{[O {\sc I}] $\lambda$ 5577\ }
\def\NaID{Na {\sc I} D\ }
\def\Msun{M$_{\rm \odot}$}
\def\qisp{Q$_{ISP}$}
\def\uisp{U$_{ISP}$}
\def\fluxu{${\rm ergs\, cm^{-2} \AA^{-1} s^{-1}}$}
\def\dm15{$\Delta M_{15}$}
\def\uisp{U$_{ISP}$}
\def\fluxu{${\rm ergs\, cm^{-2} \AA^{-1} s^{-1}}$}
\def\dm15{$\Delta M_{15}$}
\newcommand\gr{$\gamma$-ray}
\newcommand\grs{$\gamma$-rays}
\newcommand\grb{$\gamma$-ray burst}
\newcommand\grbs{$\gamma$-ray bursts}
\def \lta {\mathrel{\vcenter
     {\hbox{$<$}\nointerlineskip\hbox{$\sim$}}}}
\def \gta {\mathrel{\vcenter
     {\hbox{$>$}\nointerlineskip\hbox{$\sim$}}}}
 \def\ni{$^{56}{\rm Ni}\ $}      
 \def\Ni{$^{56}{\rm Ni}$}      
 \def\co{$^{56}{\rm Co}\ $}      
 \def\Co{$^{56}{\rm Co}$}      
 \def\fe{$^{56}{\rm Fe}\ $}      
\def\apj{Ap. J.}
\def\apjl{Ap. J. Lett.}
\def\apjs{Ap. J. Supp.}
\def\aj{Astron. J.}
\def\nat{Nature}
\def\mnras{Mon. Not. Royal. Soc.}
\def\aap{Astron. \& Astroph.}
\def\pasp{Pub. Astr. Soc. Pac.}
\def\apss{Ap. and Space. Sci.}
\def\iaucirc{IAU Circ.}
\def\araa{Ann. Rev. Astron. Ap.}

\jname{Annu. Rev. Astron. Astrophys.}
\jyear{2008}
\jvol{1}
\ARinfo{0000-0000/00/0000-00}

\title{Spectropolarimetry of Supernovae}

\markboth{Lifan Wang, \& J. Craig Wheeler (WW)}
{Spectropolarimetry of Supernovae (WW)}

\author{
Lifan Wang
\affiliation{Texas A\&M University: wang@physics.tamu.edu}
J. Craig Wheeler
\affiliation{The University of Texas at Austin: wheel@astro.as.utexas.edu}
}


\begin{abstract}
Overwhelming evidence has accumulated in recent years that 
supernova explosions are intrinsically 3-dimensional 
phenomena with significant departures from spherical symmetry. 
We review the evidence derived from spectropolarimetry that
has established several key results: virtually all supernovae
are significantly aspherical near maxim light; core-collapse 
supernovae behave differently than thermonuclear (Type Ia)
supernovae; the asphericity of core-collapse supernovae
is stronger in the inner layers showing that the explosion
process itself is strongly aspherical; core-collapse supernovae
tend to establish a preferred direction of asymmetry; 
the asphericity is stronger in the outer layers of thermonuclear
supernovae providing constraints on the burning process. We 
emphasize the utility of the Q/U plane as a diagnostic tool 
and revisit SN~1987A and SN~1993J in a contemporary context. 
An axially-symmetric geometry can explain many basic features 
of core-collapse supernovae, but significant departures from 
axial symmetry are needed to explain most events. We introduce
a spectropolarimetry type to classify the range of behavior
observed in polarized supernovae.  Understanding asymmetries in 
supernovae is important for phenomena as diverse as the origins 
of $\gamma$-ray bursts and the cosmological applications of 
Type Ia supernovae in studies of the dark energy content 
of the universe.

\end{abstract}

\maketitle

\section{Introduction}

Supernovae have been studied with modern scientific methods for 
nearly a century.  During this time, it has been traditional to 
assume that these catastrophic stellar explosions are, for all 
practical purposes, spherically symmetric.  There was 
no commanding observational need to abandon that simplifying 
assumption. Now there is.

This article will group supernovae into two major groups: 
Type Ia supernovae and all the other types that include for 
example Type II, Type IIb, Type Ib, and Type Ic.
(Harkness \& Wheeler 1990; Filippenko 1997) 
[MARGIN COMMENT Type II supernovae show distinct lines of
hydrogen. Type I supernovae do not. SN~1987A was a Type II, 
but the progenitor was a blue supergiant rather than the
more common red supergiant. Type Ia supernovae result from
exploding white dwarfs, Type Ib and Ic supernovae from
core collapse. The progenitors of Type IIb supernovae
have a thin hydrogen envelope, $\sim 0.1$\Msun.]
Evidence that supernovae may depart a little or even 
drastically from spherical symmetry has been growing for years. 
For core collapse, we know that pulsars are somehow ``kicked" at 
birth in a manner that requires a departure from both spherical 
and up/down symmetry (Lyne \& Lorimer 1994).  
The progenitor star of SN~1987A was asymmetric as are its 
surroundings and debris (Wampler et al. 1990; Crotts et al. 1989;
Burrows et al. 1995; Wang et al. 2002). The supernova remnant 
Cassiopeia A shows signs 
of a jet and counterjet that have punched holes in the expanding 
shell of debris, and there are numerous other asymmetric supernova 
remnants (Fesen, 2001; Hwang et al. 2004; Wheeler, Maund \& Couch 2008). 
Each of these observations is well known.  The question has 
been: are they merely incidental or a vital clue to how 
core-collapse supernovae work? 

For Type Ia supernovae (Hillebrandt \& Niemeyer 2000)
the progenitor white dwarf has long been treated as basically 
spherically symmetric even though the popular model is that 
the explosion must take place in a binary system where the 
white dwarf grows to the critical mass by accretion of mass 
and, inevitably, angular momentum. There could be asymmetries 
in this sort of supernova resulting from the the dynamics 
of the combustion process, the spin of the white dwarf, 
the stellar orbital motion, a surrounding accretion disk,
or the presence of the companion star. As for the case of 
core-collapse supernovae, this was known, but there was no 
compelling observational reason to consider departures from
spherical symmetry. This, too, is changing.

Our understanding of the shape of supernovae has undergone a 
revolution in the last decade.  The driving force has been a 
new type of observation: spectropolarimetry.  When light 
scatters through the expanding debris of a supernova, it 
retains information about the orientation of the scattering layers.
Since we cannot spatially resolve the average extragalactic 
supernova through direct imaging, polarization is the most powerful 
tool we have to 
judge the morphology of the ejecta. Spectropolarimetry measures 
both the overall shape of the emitting region and the 
shape of regions composed of particular chemical elements.
For a supernova with a typical photospheric radius of $10^{15}$ cm,
the effective spatial resolution attained by polarimetry at
10 Mpc is 10 microarcsec. This is a factor of 100 better 
resolution than optical interferometers can give us at a 
small fraction of their cost.  

The maturation of spectropolarimetry of supernovae over the
last decade occurred in parallel with two other immense
revolutions: the supernova/gamma-ray burst 
connection and the use of Type Ia supernovae to 
discover the acceleration of the Universe. Understanding
that core-collapse supernovae were routinely aspherical
developed in parallel with the understanding that long/soft
gamma-ray bursts resulted from tightly collimated beams
of gamma-rays from some varieties of Type Ic supernovae
(Woosley \& Blooom 2006).
Surely there is a close connection between the asymmetry
of ``normal" core-collapse supernovae and those that
yield gamma-ray bursts. For Type Ia, the knowledge that
they are asymmetric is a challenge for their ever
more precise use as cosmological tools. 
 
\section{History}
\label{history}

\subsection{Predictions of Supernovae Polarization}

Wolstencroft \& Kemp (1972) argued that an intrinsic magnetic
field dispersed by a supernova explosion might induce optical 
circular polarization.  Shakhovskoi (1976) proposed that electron 
scattering in an asymmetric, expanding envelope was the 
most likely source of any intrinsic polarization.
Shapiro \& Sutherland (1982) did the first quantitative 
study of polarimetry as an important tool to study the 
geometry of supernovae.  They found that a mixture of 
scattering and absorption can yield a significantly higher 
degree of polarization than expected from the classical, 
pure scattering atmosphere discussed by Chandrasekhar (1960). 
More sophisticated models (\ho\ 1991) predicted that, for
a given axis ratio, the extended atmospheres expected for 
supernovae would provide even more polarization than 
a plane-parallel atmosphere.

McCall (1984) argued that lines that form P-Cygni scattering 
profiles in expanding atmospheres with non-circular isophotes 
should show a linar polarization greater than that in the
continuum. He predicted that the polarization should increase 
at the absorption minimum as the absorbing material blocked
predominantly unpolarized forward scattered flux from the 
portion of the photosphere along the line of sight. This would
increase the relative proportion of polarized flux that
scatterered from the asymmetric limb.  There is also a tendency 
for the polarization to decrease at the emission peak because 
the emitted unpolarized flux will tend to dilute the underlying 
polarized continuum flux. The result has been characterized as 
an ``inverse P-Cygni" polarization profile across an unblended 
P-Cygni line in the flux spectrum.
[MARGIN COMMENT P-Cygni line profiles form when outflow
from a source, the star P-Cygni or a supernova, yields an
expanding atmosphere such that there is emission at the
rest wavelength of a line from the transverse flow and
blue-shifted absorption where the flow expanding toward
the observer obscurs the photosphere.] 

\subsection{Early Observational Attempts}

The first reported supernova polarization observations were 
by Serkowski (1970). Other early attempts were by 
Shakhovskoi \& Efimov (1972), Wolstencroft \& Kemp (1972), 
Lee et al. (1972), and Shakhovskoi (1976).
It is not clear that any of these data represented anything but
interstellar polarization. 

McCall et al. (1984) presented the first spectropolarimetric 
data on a supernova, the Type Ia SN 1983G in NGC 4753 near 
maximum light.  No change in polarization through the P Cygni 
lines was observed. The polarization at about 2 \% was probably 
due to the interstellar polarization (see \S 3.3). McCall et al. 
also looked for circular polarization, but none was detected. 
McCall (1985) reported observations of the Type Ib SN 1983n 
close to maximum light and noted that the polarization dipped 
from about 1.4 \% in the continuum to about 0.8 \% in the 
Fe II feature with no change in position angle, strongly 
suggesting an intrinsic polarization.  Unfortunately, 
the data were never published.
Spyromilio and Bailey (1993) obtained spectropolarimetry of the 
Type Ia SN 1992A 2 weeks and 7 weeks after maximum light. 
They observed no significant variations across the
spectral features. 

The first events for which systematic spectropolarimetry was 
obtained were SN~1987A, an explosion in a rare blue super 
giant, and SN~1993J, an explosion in a star nearly stripped
of its hydrogen envelope. We present a detailed discussion 
of these events below.  These two events illustrated how poor 
the overall data base of supernova spectropolarimetry was up 
through the early 1990s.

In 1994 we began a program to obtain spectropolarimetry of 
as many supernovae as possible that were visible from McDonald 
Observatory. At the time, as illustrated by this brief history, 
only a handful of events had been examined at all and there 
were virtually no statistics.  The first few supernovae 
our group studied (and those in the previous sparse record 
like SN~1987A and SN~1993J) were classified as ``peculiar" 
in some way, so we did not know whether we were seeing 
incidental peculiarities or something truly significant.

As data accumulated, however, this uncertainty was removed, 
and significant new insights were revealed. With more data 
and better statistics, we identified the first key trend:. 
the data were bi-modal (Wang et al. 1996). Type Ia 
supernovae showed little or no polarization signal near and 
after maximum light (we discuss the prominent pre-maximum 
polarization below).  Supernovae thought to arise by core collapse 
in massive stars were, by contrast, all significantly polarized. 
So far every core-collapse supernova for which we or other groups 
have obtained adequate data has been substantially polarized.  
Core-collapse supernovae are definitely not spherically symmetric. 
Table 1 gives a list of supernovae observed with photometric or 
spectroscopic polarimetry.

\section{Introduction to Supernova Spectropolarimetry}

\subsection{Stokes Parameters}
\label{stokes}

The polarization of an incoming signal can be characterized 
by its Stokes vector, S, that comprises four components, 
or Stokes parameters, I, Q, U, and V (Chandrasekhar 1960), 
where I is the intensity and Q and U measure the linear 
polarization. The component, V, measures the circular 
polarization.  While a portion of the light from a supernova 
may be circularly polarized, low flux levels have precluded 
any serious attempt to measure circular polarization, and it 
will not be considered here. The Stokes vector has an amplitude 
and a direction, but since it is related to the intensity, 
which is the square of the amplitude of the electric vector, it 
is a quasi-vector for which the directions 0\degree\ and 
180\degree\ are identical.  

In general, I is the total flux, \^Q and \^U are  
differences in flux with the electric vector oscillating 
in two orthogonal directions on the sky, with 
\^U representing angles on the sky that are rotated by 45\degree\ 
with respect to those sampled by \^Q. Throughout this review, the
normalized Stokes parameters are defined as Q\ =\ \^Q/I and 
U\ =\ \^U /I. In this way the degree of linear polarization, 
$P$, and the polarization angle, $\theta$, are expressed 
in terms of the Stokes parameters as:

\begin{equation}
P=\frac{\sqrt{{\hat Q}^2+{\hat U}^2}}{I} = \sqrt{Q^2+U^2}, 
\end{equation}

and 

\begin{equation}
\theta = \frac{1}{2} {\rm arctan}\frac{\hat U}{\hat Q},
\end{equation}

or 
\begin{equation}
Q = P\cos{2\theta}, \ U = P\sin{2\theta}. 
\end{equation}

The astronomical convention is that $\theta \ = \ 0^{\rm o}$ 
points to the North on the sky.

\subsection{Observational Techniques}

The basic techniques of spectropolarimetry are presented by 
Miller \& Goodrich (1990), Goodrich (1991), del Toro Iniesta (2003), 
and Patat \& Romaniello (2006) (see also van de Hulst 1957).
Because polarimetry involves the difference of the ordinary 
and extraordinary light rays, the per pixel errors are 
exaggerated compared to those in the total flux spectra
and are artificial because each resolution
element contains more than one pixel. It is therefore useful 
to rebin the data into bin sizes comparable to the spectral 
resolution to reduce the artificial per pixel errors and error
correlations. Care must also be taken that estimates of 
polarization from Equation 1 are not biased to large values
by noise (Simmons \& Stewart 1985; Wang, Wheeler \& \ho\ 1997).  
For a detailed discussion of procedures involved in 
reducing supernova spectropolarimetry data, see Wang, Wheeler
\& \ho\ (1997), Appendix 1 in Leonard et al. (2001), Leonard 
\& Filippenko (2001), the Appendix in Leonard et al. (2002a), 
Wang et al. (2003a), Patat \& Romaniello (2006), Maund et al. 
(2007a) and Maund (2008).

\subsection{Interstellar Polarization (ISP)}

Polarization introduced by interstellar dust in either the host
galaxy or the Milky Way Galaxy can complicate the interpretation 
of observed polarization. Fortunately, the wavelength dependence 
of the ISP is well-quantified by observations of Galactic stars
(Serkowski et al. 1995; Whitett et al. 1992). The polarization 
from supernova ejecta is expected to vary across spectral lines 
and the wavelength dependence may be different. The supernova
polarization also varies in time whereas the ISP does not. In
principle, these features allow separation of the supernova 
polarization from the contributions of the ISP. 

There is no totally satisfactory method to derive the ISP of 
supernova polarimetry observationally. Many methods have been proposed, 
but they have to make assumption about the supernova polarization. 
Examples of these assumptions are: (1) Certain components of the 
supernova spectrum, such as the emission peaks of P-Cygni profiles, 
are not polarized (Trammell et al. 1993; Tran et al. 1997); (2) 
That the blue end of the supernova spectra is unpolarized because 
of strong line blended depolarizing lines (Wang et al. 2001). The 
first method also implicitly assumes a simple axially-symmetric 
geometry of the ejecta which is often ruled out by the observational data.
It is thus often dangerous to derive scientific results assuming that
one has an accurate {\it a priori} knowledge of the (ISP).

If the ISP is significantly larger than the intrinsic polarization of 
a supernova, the wavelength dependence of the observed polarization 
can provide estimates of the properties of the dust particles along 
the line of sight to the supernova. An estimate of the wavelength at 
which the ISP reaches maximum can be compared to the that for the Galaxy, 
around 5500\AA\, to see if the mean properties of the host dust are 
similar or different.  The ISP due to the dust lanes in Centaurus~A 
was probed by observations of SN~1986G and it was concluded that
the size of dust particles are smaller than typical Galactic dust
particles (Hough et al. 1987). Recently Wang et al. (2003a) studied
the dust properties in NGC 1448 and found the size of  dust particles
are slightly smaller than their Galactic counterparts, but are
nonetheless consist with the properties observed in the Galaxy.
Leonard et al. (2002b) observed a highly reddened Type II plateau
supernova (see \S \ref{SNIIP}) SN~1999gi in NGC 3184, and found that 
the polarization efficiency of the dust in NGC 3184 to be much higher 
than typical Galactic dust. It is certainly important to get more 
polarization data of highly extinct supernovae as this is a powerful 
tool to study the elusive properties of extragalactic dust.

\subsection{The Dominant Axis}
\label{dominant}

The prominent spectral features of supernovae make it possible 
to deduce valuable information without accurate estimates of the ISP.  
Wang et al. (2003a) outlined a method to decompose the observed 
polarimetry into two components, equivalent to a principle component 
analysis with two components. In the Q/U plane, the two components 
correspond to the polarized vectors projected onto the so-called 
{\it dominant axis} and onto the axis perpendicular to the dominant axis, 
which we refer to as the {\it orthogonal axis}. A dominant axis can often 
-- but not always -- be defined from the distribution of the data points 
on the Q/U plane (Wang et al. 2001).  The spectropolarimetry projected to 
the dominant axis represents global geometric deviations from spherical 
symmetry, whereas the vector perpendicular to the dominant axis
represents physical deviations from the dominant axis. The 
coordinates of the new system are given by rotating the original 
coordinates counterclockwise so that the Q axis overlaps the dominant 
axis in the new coordinate system and the dominant axis points toward 
the center of the data cluster on the Q/U plane. The components parallel 
and perpendicular to the dominant axis are given by 
\begin{equation}
\label{dom}
P_d\ = \ (Q-Q_{ISP})\cos\alpha\ + \ (U-U_{ISP})\sin\alpha,
\end{equation}
and
\begin{equation}
\label{orth}
P_0\ = \ -(Q-Q_{ISP})\sin\alpha\ + \ (U-U_{ISP})\cos\alpha,
\end{equation}
where $P_d$ and $P_0$ are the polarization components parallel to 
the dominant axis and orthogonal to that axis, respectively, 
$Q_{ISP}$ and $U_{ISP} $ are the Stokes parameters of the ISP, 
$\alpha\ = 2\theta_d$ is the rotation angle in the Q/U plane and 
$\theta_d$ is the polarization position angle of the dominant axis. An 
advantage of using these decomposition formulae rather than
calculating the degree of polarization as given by Equations 1 and 2 
is that the spectral profiles of $P_0$ and $P_d$ are insensitive 
to the choice of ISP. A similar procedure producing ``rotated
Stokes parameters" was employed by Leonard et al. (2001) who 
calculated a fit to the continuum position angle so that all 
the continuum polarization falls on the rotated Q. This is 
equivalent to, but not quite the same technically, as fitting a 
dominant axis in the Q/U plane. The polarization spectral profiles 
of the latter decomposition method do not depend on the unknown ISP.

If the supernova ejecta are smooth and axially symmetric, 
the data points on the Q/U diagram should follow a straight line
defining the dominant axis, as expected from theoretical models 
(H\"oflich et al. 1996). If the geometry departs significantly 
from axial symmetry, the data points will show noticeable 
dispersion from a straight line, in the direction of the
orthogonal axis.  These basic situations and their representation 
in the Q/U plane are shown schematically in Figure~1.

In the top left of Figure~1, we show a smooth, 
axially symmetric geometric structure with 
the axis tilted at an arbitrary position angle on the sky
and, in general, with respect to the line of sight.
For homologous expansion, v $\propto$ r, surfaces
of constant velocity are planes normal to the line of sight.
Viewing at different wavelengths through a line profile corresponds
to viewing the geometry along different slices normal to
the line of sight. In general, due to the interplay of 
geometry and optical depth, different wavelengths will
register different amplitudes of polarization and hence
values of Q and U, but at a fixed angle. The resulting 
wavelength-dependent polarization amplitude plotted in 
the Q/U plane will follow a straight line (top right), 
the dominant axis. A simple rotation of the coordinates 
would align the rotated Q axis with the line. In this simple 
example, there would be no contribution along the 
orthogonal axis corresponding to the rotated U axis. 

In the lower left of Figure~1 we schematically illustrate
a case for which the axisymmetry is broken, in this case
by ``clumps" of composition-dependent and hence optical
depth- and wavelength-dependent structure. In practice,
clumps of high-opacity, absorbing material will block
parts of the underlying photosphere. This will induce
wavelength-dependent geometry on constant velocity slices
even if the underlying photosphere is symmetric and hence
imposes no net polarization itself. Clumps blocking
an asymmetric photosphere will yield more complex structure,
but no difference in principle.  In this case, the polarization 
distribution in the Q/U plane will no longer be along a straight 
line (lower right). The basic axisymmetric geometry may
still be evident in terms of a dominant axis, as illustrated, but 
the departure from axial symmetry caused by the ``clumping" will 
yield a finite, and physically significant, distribution along 
the orthogonal axis. We use this basic phenomenology to define
spectropolarimetry types in \S \ref{types}. 

The decomposition of the polarization components given by 
Equations \ref{dom} and \ref{orth} assumes the existence of only 
two components with fixed axes: an ISP component and only 
one additional component that is intrinsic to the supernova. 
This is clearly an oversimplification for some events,
for which a more complicated decomposition is needed. A more
realistic approach is to assume that the line polarization and 
continuum polarization have independent geometries and hence that 
the observed polarization is a combinations of the two.

While there are numerous ways of presenting and analyzing 
spectropolarimetry data, one clear advantage of displaying the data 
in the Q/U plane is that it facilitates doing the requisite ``vector 
analysis" by eye. One can immediately and directly see how the
amplitude and polarization angle will change with a change in the 
placement of the always uncertain and frequently controversial ISP. 
Likewise, the tendency of the data to fall along a single axis, or 
to depart from a primary axis in loops or other structures, is 
transparent, completely independent of the placement of the ISP. 
We will take advantage of this power in the subsequent analysis.

\subsection{Loops in the Q/U Plane}
\label{loops}

In general, the continuum polarization from a supernova provides 
information about the overall shape of the photosphere, whereas 
the polarization across spectral lines is more sensitive to 
small-scale structures in the ejecta. Large-scale gradients in 
density and composition structure may also affect 
line polarization, but such effects are generally saturated 
for strong lines due to their large optical depth.  

If the composition structure is complex, then the degree and 
angle of polarization may vary significantly across strong spectral 
lines.  The result is to produce ``loops" in the Q/U plane. 
These loops are loci on the Q/U plane that are functions of
the wavelength across the line feature and hence functions of 
the velocity and of the depth of the portion of the structure
that contributes to that wavelength. Because the loops
represent changes in the amplitude and angle of polarization
as a function of wavelength, velocity, and depth, they are
specifically evidence for breakdown of axial symmetry. The 
loops could be caused by composition-dependent clumps or more 
organized structures that nevertheless break the axial symmetry. 

An example of how loops may be formed is given in Figure 2 
taken from Kasen et al. (2003), a paper that also has an excellent 
pedagogical exposition of polarization in supernovae atmospheres.  
Figure 2 was constructed in the context of the high-velocity
calcium feature of Type Ia supernovae (see \S \ref{hiVCa}), but the
principles are quite general. In the case illustrated, 
an edge-on torus surrounds an ellipsoidal 
photosphere, both expanding homologously. That geometry breaks 
the axisymmetry.  The torus has a strong absorption line that 
blocks the continuum emission from the photosphere, but the
geometry of the blocking region depends on the velocity
slices through the structure and hence on the wavelength
across the line profile. The resulting locus in the Q/U plane 
shows a loop that matches the observations (of Type Ia SN~2001el)
rather well. Complications that can be envisaged,
many of which are discussed by Kasen et al., are an asymmetric
photosphere, different geometries for the surrounding shell,
and breaking up of the structure into clumps. 

As we shall see in the subsequent discussion, loops in the
Q/U plane, with their implied non-axisymmetric structure, are 
common in both core-collapse and thermonuclear supernovae.
These loops represent an important new clue to the physics and
the structure of both types of explosion.

\subsection{Spectropolarimetry Types}
\label{types}

We can classify the spectropolarization characteristics of a 
supernova according to its distribution in the Q-U diagram.  
This classification scheme follows directly from the 
principles outlined in Figures~1 and 2 and is insensitive to the exact 
value of the ISP. We define the following spectropolarimetric
(SP) types: 

{\it SP Type N0:} This type is characterized by polarization data with 
no significant deviation from a distribution that is consistent with 
observational noise. The centroid of the distribution 
may be offset from zero polarization due to interstellar dust, but
there is no measurable supernova polarization. There is no dominant axis.

{\it SP Type N1:} This type is characterized by polarization data with 
no significant elongation in any preferred direction, but for which 
the data distribution is wider than would be consistent with the 
observational errors, indicating intrinsic polarization or systematic 
underestimate of observational error. There is no dominant axis.

{\it SP Type D0:}  This type is characterized by data that show an 
elongated ellipse in a Q/U diagram, with the distribution orthogonal 
to the major axis of the ellipse consistent with observational noise. 
The locus of the data can be well approximated by fitting a straight 
line (except in the case of very strong interstellar polarization and 
very broad wavelength coverage). The position angle of this fitted
line defines the dominant axis of the data.

{\it SP Type D1:} This type is characterized by data that show an
elongated ellipse in a Q/U diagram so that a dominant axis can be 
identified, but for which a straight line does not provide a satisfactory 
fit. Significant deviations are found orthogonal to the dominant axis.

{\it SP Type L:} Some supernovae show large significant changes in the
amplitude and position angle across strong spectral lines. These variations 
results in prominent loops on the Q/U diagram. 

In general, the SP type is deduced from observations 
at a certain epoch and in a certain wavelength range. 
The spectropolarimetric properties of supernovae evolve with time, so 
the SP type as defined here can evolve as well.
In addition, different portions of the spectrum may reveal different
characteristics. The continuum or a single line could be a D0, 
while another line could be an L. We have assigned tentative
SP types to those supernovae in Table 1 where
the data justified it. We also use this classification to
discuss individual events in the remainder of this review.

In order to characterize the data, it is critical to correctly 
estimate both the systematic and the statistical observational 
errors. In additional to normal spectropolarimetric 
calibrations such as from observations of polarized and 
unpolarized standard stars, it is highly advisable to acquire 
two complete sets of polarization data each night. This allows 
the errors and stability of the spectropolarimetry to be cross 
checked with data taken with identical observational setups. 

\section{Core Collapse Supernovae}

\subsection{General Trends - Evidence for Bipolar Explosions in the Machine}
\label{bipolar}

The single most important conclusion to arise from the
study of supernovae spectropolarimetry is that the process
of core collapse routinely involves an intrinsically
strongly aspherical explosion mechanism that is directed
substantially along a single direction in space. Prior to
these spectropolarimetic studies, the explosion was assumed to be 
substantially spherical with any departures from spherical symmetry
being due to rather small scale effects of convection
or Rayleigh-Taylor instabilities. The evidence for
strongly directed explosions has given rise to a new
literature of ``jet-induced" explosions and variations
on that theme (Khokhlov et al. 1999; MacFadyen \& Woosley, 1999; 
Wheeler et al. 2000; Wheeler, Meier \& Wilson 2002; 
Shibata et al. 2003; Maeda \& Nomoto 2003; Kotake et al. 2004; 
Yamada \& Sawai 2004; Kifonidis et al. 2006; Obergaulinger et al. 2006;
Burrows et al. 2006, 2007; 
Mezzacappa \& Blondin 2007; Moiseenko \& Bisnovatyi-Kogan 2007;
Komissarov \& Barkov 2007; and references therein) 
and established a strong tie to cosmic
long gamma-ray bursts that also involve the collapse
of massive stars and highly directed energy output
(Woosley \& Bloom 2006). Major efforts are now 
underway to understand the origin of the asymmetry,
most plausibly in terms of rotation and magnetic
field in the progenitor and newly-formed neutron star
(Akiyama \& Wheeler 2003; Thompson, Chang \& Quataert 2004;
Masada et al. 2006; Udzenski \& MacFadyen 2007; and references
therein). The spectropolarimetry of core-collapse supernovae has 
thus helped to catalyze a paradigm shift in thinking 
about the explosion mechanism. 

We now recognize that a dominant axis in the polarimetry 
represents a primarily axisymmetric activity. 
In the subsequent sections we will summarize in some detail how
the data that led to this conclusion were attained, re-visit
SN~1987A and SN~1993J in the current context, and attempt to 
establish a common framework for the interpretation of the data.
We will establish that polarization is not a feature of a few
odd core-collapse supernovae, but a generic feature of every 
spectral type. A recent key conclusion is that while a ``bipolar" 
structure is dominant in the geometry of core-collapse supernovae, 
departures from axisymmetry evidenced by loops in the Q/U plane 
and other composition-dependent structures are also ubiquitous 
and must be incorporated in physical models.

\subsection{SN~1987A}
\label{sn1987A}

\subsubsection{Photometric Polarimetry}

SN~1987A was the first supernova for which good photometric and 
spectropolarimetric data were obtained. Although formally a
hydrogen-rich Type II supernova, the blue supergiant progenitor
and detailed observations of all kinds (Arnett, Bahcall,
Kirshner \& Woosley 1989; McCray 1993) make this supernova
a category unto itself. The polarimetric data on SN~1987A have not 
been adequately explored. We can view them now in the context of
the larger sample of more distant supernovae. The clear
conclusion is that more thorough quantitative analysis is warrented,
but the basic themes of core-collapse supernovae are reflected
here. SN~1987A displayed significant large-scale asymmetry with 
well-defined principle axes consistent with jet-like flow, but 
was also marked by departures from axisymmetry. For a complete
set of references, see Jeffery (1991a) and Wang et al. (2002).

Some of the early estimates of ISP now appear to have been incorrect and
to have led to incorrect conclusions.
With the later and probably more reliable
estimate of the ISP, Jeffery (1991a) found that
the polarization was small at early
times and then grew monotonically for the first month.
SN~1987A gave a hint that the inner machine of the explosion was
strongly asymmetric, evidence that has proven ubiquitous for core
collapse with current, systematic observations.
Theoretical models to account for the early behavior of SN~1987A
were presented by M\'endez et al. (1988), Jeffery (1989), Jeffery
(1990), Jeffery (1991b) and H\"oflich (1991). Models based on the estimate
of the ISP by M\'endez et al. that implied an early declining value of
the polarization need to be reconsidered.

We reproduce in Figure 3 a portion of the data presented by Jeffery 
(1991a). Figure 3
suggests that around day 140, as the major light curve maximum 
gave way to the exponential radioactive tail, the polarization 
jumped to 1.3 -- 1.5 \% and then slowly dropped back to around 
0.2 -- 0.4 \% by day 200. This jump might be associated with the 
photosphere receding through the outer hydrogen envelope and 
revealing the inner core. 
Whether the polarization of SN~1987A was quite 
variable in this epoch or some of the data are questionable has 
yet to be resolved, calling for more study and careful modeling. 

A remarkable fact is that the broad-band polarization angle 
did not waver through this whole evolution, including the possible
large spike in the broad-band polarization.  
The second speckle source, the ``mystery spot," 
[MARGIN COMMENT The ``mystery spot associated with SN~1987A 
(Meikle et al. 1987; Dotani et al. 1987;
Nisenson \& Papaliolios 1999) was
a small source of luminosity detected in speckle imaging that
was nearly as bright as the supernova. No thoroughly satisfactory
explanation has been provided.]
and the orientation of the late-time HST image of the ejecta
at $\theta \sim 16$\degree\  provide another measure of the
orientation (Wang et al. 2002). For compactness, we will refer to this as the 
``speckle angle" in the subsequent discussion. Scattering from features
aligned in this direction would give a plane of polarization with
position angle of 106\degree. Within observational uncertainty,
this is very close to the position angle of the continuum
polarization.  The position angle of the minor axes of the 
circumstellar rings is also aligned close to the position angle 
determined from the dominant axis of polarization. 

By these measures, SN 1987A ``pointed" in a certain direction, 
a position angle of $\theta \sim$ 16\degree, and maintained that 
orientation throughout its development (Wang et al. 2002). This 
requires a large-scale, systematic, directed asymmetry that cannot be 
accounted for by small-scale convection or turbulence. 
The continuum polarization probes the geometric structure of
the photosphere.  The persistent polarization position angle of 
the dominant axis across a broad wavelength range, and the lack of 
significant evolution of the polarization position angle strongly 
argue that the geometry of the photosphere at early and late epochs 
share the same geometric structure. The data also show that the 
overall structure of SN~1987A was remarkably axially symmetric from 
deep inside the oxygen rich zone out to the hydrogen envelope
and on to the circumstellar rings.  The totality of the data suggests 
that the angular momentum of the progenitor system played a 
crucial role in the explosion of SN~1987A (Wang et al. 2002).  

\subsubsection{Spectropolarimetry}

The first spectropolarimetry of SN~1987A, obtained on 8 March, 1987,
was presented by Schwarz \& Mundt (1987) and modeled by 
Jeffery (1987).  Broadband polarimetry was also reported by 
Clocchiatti \& Marraco (1988).  Subsequent excellent spectropolarimetry of 
Cropper et al.  (1988) confirmed that as data are tracked 
as a function of wavelength over spectral features, the 
polarization angle does sometimes change with wavelength, 
giving rise to ``loops" in the Q/U plane. 
As explained in \S \ref{loops}, as wavelength varies
over a Doppler-broadened line feature, different depths are
sampled in the ejecta, with a given wavelength sampling
a velocity ``slice" through the ejecta normal to the 
line of sight. The loops imply that the polarization amplitude 
and angle are varying along those velocity/depth slices.   
This means that there must be some substantial departure 
from axisymmetry in SN~1987A imposed on the overall ``pointed" 
behavior revealed by the photometric polarimetry. As we will
now illustrate, this spectropolarimetric behavior gives a 
rich phenomenology that is ripe for progress in physical 
understanding of SN~1987A. These loops will give greater 
insight into the composition-dependent three-dimensional 
structure of the ejecta and are worth study in much more depth 
than allowed here. In subsequent sections, we will show
that these loops and the implied non-axisymmetry, are not
unique to SN~1987A, but common to all core collapse events.

We describe here some of the data of Cropper et al. (1988) 
corrected for the ISP using the data of M\'endez (1990) as 
parametrized by Jeffery (1991a).  As shown in the Q/U diagram
of Figure 4 (left panel), 
when the supernova was 
about 12 days old, the data in the vicinity of the H$\alpha$ 
line extend to negative U values corresponding to a position angle 
$\theta \sim 150$\degree\ with respect to the polarized continuum
adjacent to the line. This is similar to, but perhaps significantly 
different than, the nearly time-independent V-band position angle 
$\theta \sim 120$\degree\ in Figure 3, 
and the speckle angle.

Figure 4 (right panel) 
also shows the dramatically different distribution of the 
H$\alpha$ polarization in the Q/U plane when the supernova was about 
70 days old and approaching maximum brightness.  Although 
the inner, asymmetric core of the star has yet to be exposed, 
the variation of polarization across the H$\alpha$ line has
changed dramatically. The variation of Q and U with wavelength 
across the H$\alpha$ line profile now traces out a distinct 
loop in the Q/U plane. For reference, the H$\alpha$ absorption 
minimum at 6460 \AA\ corresponds to the maximum positive Q and U.  
The distribution at this epoch has little relation to 
the orientation displayed at the earlier epoch in the left-hand 
panel of Figure 4.  One significant new feature is the distinct loop 
of green points that fall very close to the speckle angle.  Although 
the photosphere is still in the hydrogen envelope, the hydrogen 
is clearly, in part, reflecting the dominant asymmetry of the 
inner regions. A likely cause is the non-spherical distribution 
of the source of ionization in the form of a ``lump" of radioactive 
nickel and cobalt (Chugai 1992). Departures from the dominant axis
lead to loops associated with H$\alpha$. All the complexities displayed 
here are topics for more in-depth study.

The next epoch presented by Cropper et al. (1988) is
on June 3, 1987, about 100 days after the explosion, midway
through the decline of the light curve to the radioactive tail
and about the time the V-band polarization jumps. A sample of
these data is shown in Figure 5. 
In the vicinity of H${\alpha}$ 
there is again a distinct extension roughly along the speckle angle, 
but also an interesting loop structure that reflects the 
non-axisymmetric interplay of the line opacity with the polarized 
continuum.  The absorption minimum at $\sim 6400$ \AA\ corresponds 
to the data of most extreme positive Q and U. Significant polarization, 
primarily along the speckle angle, is shown by H$\alpha$ up to a year 
after the explosion (Cropper et al. 1988).

The He I $\lambda$5876 line and surrounding continuum 
illustrated in the right panel of Figure 5
show remarkable uniformity. The data span Q = 0 at U $\sim -1$ with 
an exceedingly well-defined dominant axis with position 
angle, $\theta \sim 4$\degree.  The displacement to negative U can be 
understood as resulting from the addition of the underlying 
continuum that adds a wavelength-independent component to the
line polarization (Wang et al.  2003b). The polarization angle of 
this helium feature is not the speckle angle. The simple 
single-axis behavior of the He line is in stark contrast with that 
of the H$\alpha$ feature.  It is curious that the distinct orientation 
of the He I line is not imprinted in some way on the H geometry. 

While most of the polarization of SN~1987A is intrinsic, the
possibility remains that some effects could be induced by
dust scattering in the asymmetric environment (Wang \& Wheeler 1997)

As in so many other ways, SN~1987A was also a canonical
event in terms of its polarimetry. The data deserve
a much more thorough quantitative study than has been
done, or than we can attempt here, but several key
themes emerge. The explosion was definately asymmetric.
There is evidence from spectropolarimetry in the continuum and across 
the H$\alpha$ line for a well-defined physical orientation 
that corresponds to other evidence for a substantially
axially-symmetric structure and ``jet-like" flow. On
the other hand, there are fascinating and substantial
departures from a simple ``bi-polar" explosion. 

From the perspective of SP types defined in \S \ref{types}, the 
data can be classified as SP  Type L with respect to strong 
P-Cygni lines: H$\alpha$, O I, and Ca II. Across the entire 
optical wavelength range, the supernova shows an apparent
dominant axis which is nearly constant with time and can be
classified as SP Type D1. The He I line in Figure 5 can be classified
as SP Type D0. 

\subsection{Type II Supernovae}
\label{SNII}

\subsubsection{Type II Plateau}
\label{SNIIP}

SN~1987A was clearly asymmetric with strong evidence for
bi-polar, jet-like flow. This supernova also had obvious
peculiarities; its progenitor was a blue supergiant and
a binary companion might have been involved. The question
remained whether the asymmetries revealed in SN~1987A were
peculiar to it, or clues to general processes in core collapse.
One means to address this is to examine other Type II supernovae. 

Type IIP supernovae explode in red supergiants with extended
hydrogen envelopes. Their light curves display a long
phase of weeks to months of nearly constant luminosity, 
called the plateau phase, as the envelope expands and
the photosphere recedes by recombination through the
envelope. These supernovae are thought to arise in single
stars of modest mass, $\sim 10 - 20$ \Msun, and hence
to be the most straight forward, ``generic" type of 
core collapse supernova. For that reason, it is especially
important to investigate their polarization properties.
Spectropolarimetry of Type~IIP was reported by 
Leonard \& Filippenko (2001) and Leonard et al. (2002b).
Other examples are explored in some detail below.

For Type IIP supernovae, the polarization 
is usually modest on the plateau phase, but probably not zero. 
The density structure of the envelope is expected to be 
nearly spherically symmetric because conservation of angular
momentum is likely to lead to slow rotation as the envelope
expands after core hydrogen depletion. A likely cause of the 
early polarization is an asymmetric distribution of radioactive 
elements that distorts the ionization and excitation structure 
even though the density structure remains essentially spherically 
symmetric (Chugai, 1992; \ho, Khokhlov \& Wang, 2001). 

One of the best studied examples of a Type II was SN~1999em
that displayed a classic, if long-lived, plateau suggesting 
an especially massive outer hydrogen envelope (Leonard et al.
2001; Wang et al. 2002b). Observations were obtained from
about a week after discovery to about a half-year later.
Data for SN~1999em on a Q/U plot are shown in Figure 6. 
The striking feature is that the data points concentrate on a 
single line as the wavelength and time vary. The same behavior 
is well-illustrated in Figure 8 of Leonard et al. (2001) that 
showed that the polarization of SN~1999em jumped from about 
0.2\% in the early phases to about 0.5\% 160 days later, but 
with virtually no change in the polarization angle. The constancy 
of the position angle suggests a well-defined symmetry axis 
throughout the ejecta and independent of wavelength. Leonard et al. 
note hints of non-axisymmetry (their Figure 16), but the data 
is insufficient for detailed study.  SN~1999em was a rather 
typical Type II plateau explosion, yet it demonstrated the 
key behavior of core-collapse supernovae emphasized in this 
review: the explosion of SN~1999em was aspherical and aligned 
along a fixed direction in space. SN~1999em would be classified
as a SP Type D0 or D1; a careful study has not been made of the 
dispersion in the orthogonal direction.

As a Type II plateau supernova continues to expand, the
photosphere finally recedes through the hydrogen envelope
exposing the ejecta from the heavy-element core. This is 
accompanied by a sharp drop in luminosity characterizing
the end of the plateau phase of the light curve. The plateau
phase is typically followed by an exponential decline in
luminosity consistent with the decreasing deposition of energy 
from radioactive decay.  The polarization tends to rise rapidly
at the end of the plateau phase as the photosphere enters the
core ejecta. This was suggested in SN~1987A, in which the 
polarization jumped from 0.2 to 1.2\% at 110 days as the maximum 
peak gave way to the radioactive decay tail. 

This sudden transition 
was clearly observed in the careful campaign on SN~2004dj by 
Leonard et al.  (2006), as illustrated in Figure 7. The black
points in the left panel give the visual brightness and show 
the end of the plateau phase and the transition to the exponential 
tail that is presumed to result from radioactive decay.
The red boxes show the dramatic rise in the continuum
polarization as the inner core is revealed. The right hand
panel in Figure 7 shows the behavior in the Q/U plane from
the spike in polarization at 91 days through the subsequent
decline as the ejecta thin out and become optically thin
to the scattering that induces the polarization. Note the
rotation of the polarization angle between the point at 91
days and the point at 128 days by about 30\degree, after
which the angle remains essentially constant. 

The origin and evolution of the polarization of SN~2004dj 
has been discussed and modeled by Chugai et al. (2005) and  
Chugai (2006) who concludes that a bipolar ejection of
radioactive $^{56}$Ni with more mass in the near rather than
the far hemisphere can account both for the polarization
seen in SN~2004dj and the asymmetric profile of the 
H$\alpha$ line observed in the nebular phase. Once again, the 
polarization of SN~2004dj is consistent with a jet-like flow 
that originates deep in the explosion itself. The rotation
of the polarization position angle may suggest that
the nickle is broken into clumps (Leonard et al. 2006), that 
other composition inhomogeneities disrupt the photosphere
(Chugai 2006) or that the the axis of the ``jet" is tilted 
with respect to another measure of the geometry in the inner 
regions, perhaps the rotational axis of the core (\S \ref{sn2005bf}). 
The answer could, of course, be a combination of these effects.

\subsubsection{Type IIn}

Type IIn supernovae are characterized by hydrogen-rich
ejecta with narrow emission lines (Schlegel 1990) that
are attributed to emission associated with substantial
circumstellar interaction. The differences in the
progenitors and their evolution that distinguish these
events from classical Type IIP is not clear. Spectropolarimetry
can provide useful new information to address the issue.

The first Type IIn with spectropolarimetry was SN~1998S.  
This event was observed by Leonard et al. (2000) about two 
weeks before maximum light and by Wang et al. (2001)
about 10 and 41 days past maximum light.  SN~1998S 
displayed narrow emission lines characteristic of 
Type IIn, but also lines characteristic of Wolf-Rayet
stars that are not usual for this class. This suggests
that SN~1998S had already lost substantial mass from
its envelope so that, while very interesting, it may
not be representative of all Type IIn. Leonard et al. (2000) 
favored the interpretation that the broad lines were 
unpolarized and the narrow lines were polarized. With the 
benefit of two more epochs of data and the assumption that 
the ISP does not vary with time, Wang et al. (2001) made a 
rather different estimate of the ISP.  This would give 
a continuum polarization at the first epoch of $\sim$ 1.6\%. 
This estimate of the ISP also implies that both the broad lines 
and the narrow lines are polarized in the pre-maximum spectrum 
with the continuum displaying an essentially constant position angle. 
Ten days after maximum, the continuum shows a somewhat smaller 
polarization with a rotation of about 34\degree\ compared to the 
pre-maximum data. The polarization then grows to about 3\% while
maintaining the same position angle. 

A full analysis of the polarization of Type IIn supernovae must also
account for the physical state and geometry of the circumstellar
medium (Chugai 2001). Hoffman et al. (2007) showed that Type IIn 
SN 1997eg displayed loops in the Q/U plane over the H$\alpha$ line. 
This implies that the continuum and line-forming regions had 
different geometries with different orientations.  Hoffman et al. 
suggest the that ejecta had an ellipsoidal photosphere and that 
the circumstellar medium where the hydrogen lines were formed had 
a flattened disk-like profile. Deeper investigation of this supernova 
and the polarization of the narrow lines associated with SN~1998S are 
warrented.

Type IIn supernova usually show a clear dominant axis. They 
are thus usually SP Type D0 or D1 in the 
continuum, Some Type IIn show loops across spectral lines
Those features would be SP Type L. 

\subsection{Type IIb: SN 1993J and Similar Events}

In the context of asking whether or not the asymmetries of 
core-collapse supernovae are peculiar to individual events or 
representative of generic phenomena and whether the
asymmetries are associated with the fundamental mechanism
of core collapse or peculiarities of the environment, 
the Type IIb supernovae represent an important test case.
[MARGINAL COMMENT The Type IIb events are defined as those that show 
distinct evidence of hydrogen at early times, but then distinct evidence 
of helium -- similar to a Type Ib -- at later epochs. These
events are deduced to have lost all but a very thin layer of 
hydrogen, probably by transfer to a companion. The binary
companion to the prototype Type IIb SN~1993J has been specifically 
identified (Maund et al. 2004).] 
By examining the Type IIb, one is 
intrinsically looking closer to the underlying machine of the 
explosion. Important issues are whether this class of events 
displays properties similar to other core-collapse supernovae or 
properties especially characteristic of their class that might 
be clues to their origin. 

\subsubsection{SN~1993J}
\label{sn1993J}

[MARGINAL COMMENT
SN~1993J showed an initial spike and decline in luminosity 
attributed to the ``fireball" phase as the shock hit the 
surface of the star and the shocked material rapidly adiabatically 
expanded and cooled. This was followed by a secondary peak and
a final exponential tail, both presumed to have been powered by 
radioactive decay. The emergence of the helium lines that mark
the class appeared near to the second peak. For an early summary 
of the polarization data, see Wheeler \& Filippenko (1996).]

Doroshenko, Efimov \& Shakhovskoi (1995) presented 6 epochs 
of UBVRI polarimetry of SN~1993J spanning the rise and decline associated 
with the second peak in the light curve. 
They found that the U and B data followed a fixed orientation in the 
Q/U plane, but that the longer wavelength data formed arcs or 
partial loops with time. Doreshenko et al. suggested that the
intrinsic polarization may contain two components that have 
different orientation and time variability. From the perspective 
of over a decade later, these remarks seem prescient. 

Spectropolarimetry was presented by Trammell, Hines \& Wheeler 
(1993) and by Tran et al. (1997). 
Both papers assumed that the intrinsic polarization
was zero at the peak of the H$\alpha$ line, but differed
on their treatment of the underlying continuum. We now
recognize that H$\alpha$ can be blended with He I 6678
making the assumption that the H$\alpha$ is unpolarized
highly questionable (Maund et al. 2007a). 
Re-examining the data employing the 
constraint that the blue end of the spectrum may be 
substantially de-polarized, the estimate of the ISP by 
Tran et al. appears to be preferable to that of Trammell et al.,
although there are questions about the polarization
angle assigned by Tran et al. that may affect interpretation of the data.
While we believe that the estimate of the ISP
by Tran et al. is preferable to that of Trammell et al.,
we also note that the method used by Tran et al. to subtract 
the continuum contribution and fit for the ISP can have 
significant uncertainties in both the amplitude and angle 
that they do not assess.  

Tran et al. note that they detect the rotation of the position
angle across line features in their latest data showing
that the emission lines of He I, Fe II, [O I], and H are all 
intrinsically polarized with position angles that are different 
from that of the continuum. This also suggests some degree of 
non-axisymmetric structure. The data for SN~1993J displayed
in a Q/U diagram show little evidence for a dominant angle,
but real deviation from zero polarization. SN~1993J is
classified as an SP Type N1.

\subsubsection{SN~2001ig}
\label{sn2001ig}

The Type IIb SN~2001ig was observed by Maund et al. (2007a)
over multiple epochs.  They deduced low intrinsic polarization 13 days 
after the explosion, $\sim$ 0.2\%. This is consistent 
with a nearly spherical hydrogen-rich envelope.
There is a sharp increase in polarization and a rotation of
the polarization angle by 40\degree\ by 31 days after 
explosion, after the hydrogen envelope has become 
optically thin.  At this later epoch, the data conform
to a dominant axis and would be SP Type D0. 
The rotation of the position angle shows that the asymmetry of the second 
epoch was not directly related to that of the first epoch. As 
illustrated in Figure 8, the most highly polarized lines in 
SN~1991bg showed wavelength-dependent loop configurations in 
the Q/U plane (\S \ref{loops}) so that these lines, especially
He I are SP Type L. This demands some non-axisymmetric 
structure.  SN~1991bg was one of the first examples of this loop 
morphology, a characteristic we have come to recognize as generic 
in core-collapse explosions. We cite other examples throughout
this review. 

The polarization properties of SN~2001ig are similar to SN~1993J 
and SN~1996cb, but not identical. The observing epochs were not
the same and the polarization evolves, so care must be taken
in making comparisons. The data suggest that Type IIb
supernovae may arise in similar binary systems, but not be so 
identical that they require identical observer angle, as suggested 
by Wang et al. (2001). A further implication is that the
class of Type IIb supernovae is not simply an artifact of
viewing angle. Spectropolarimetry may thus address an important
open issue and invites deeper exploration of the nature of the
evolution that leads to the progenitors of Type IIb explosions 
rather than to other types of hydrogen-stripped events. One
possibility is that mass transfer in a binary system does
not strip the whole hydrogen envelope and that some other
process, perhaps a strong stellar wind, is necessary to produce
progenitors that are completely stripped of their hydrogen. 
We explore those sorts of events in the next section.

\subsection{Stripped Core (Type Ib and Ic) Supernovae}

Type Ib supernovae are classified as those that show strong
evidence for helium lines, but scarcely any evidence for 
hydrogen in the total flux spectra. The total flux spectra
of Type Ic supernovae show little evidence for either hydrogen
or helium. Among the open issues are why Type Ib and Ic are
more devoid of hydrogen than Type IIb, how Type Ic are 
able to lose much of their helium as well, and which
Type Ic are connected to gamma-ray bursts.  A goal of continued 
study of Type Ib and Type Ic supernovae is to seek deeper 
understanding of the properties of asymmetry that are common to 
all core collapse events and hence important clues to the 
physical mechanism of the explosion.  Spectropolarimetry may also 
help to better understand why some progenitors expell virtually 
all their hydrogen and others, the Type IIb, retain a small amount.

Types Ib and Ic supernovae happen in stars that have already 
shed nearly all of their outer hydrogen layers, so they 
allow us to see deeper into the heart of the exploding star. The
trend that these events tend to show more polarization than
Type II supernovae was part of our argument (\S\ref{bipolar}, 
Wang et al. 2001) that it was the machinery of the explosion 
that was aspherical. In this section we explore the data that 
gave rise to that key conclusion. 

\subsubsection{SN~1997X}
\label{sn1997X}

The first event that provided hints that Type Ic were more 
polarized than Type II and hence that the machinery of core 
collapse was intrinsically strongly asymmetric was the Type Ic 
SN~1997X (Wang et al. 2001). SN~1997X showed exceptionally high 
polarization, as much as 7\%. Much of this was due to ISP, but
the time-dependence showed that there was substantial intrinsic
polarization.  Wang et al. suggested that the polarization
of SN~1997X may have been as high as 4\%. There seems to
have been a steep drop in polarization in the first 10 days
or so, a factor that requires explanation.

SN~1997X displayed little or no evidence of helium
lines in its total flux spectrum. That was the basis for its
classification as a Type Ic. As illustrated in Figure 9, however, 
the polarization spectra showed clear spectral features 
associated with He I $\lambda$5876 and $\lambda$ 6678. 
The helium absorption features 
peak at 15,000 \kms\ and the wings reach speeds of 28,000 \kms. 
This again shows that spectropolarimetry is a useful tool to reveal 
important aspects of the progenitor evolution by more tightly 
constraining the amount and geometry of the helium left in 
the outer layers of the progenitor. The spectropolarimetry
supports arguments that there is a continuum between the Type Ib 
and the Type Ic, rather than a distinct transition between
the two spectral classes. In the Q/U plane, SN~1997X evolves
from an SP Type N1 around maximum to an SP Type  D1 about a month later.

\subsubsection{SN~2005bf}
\label{sn2005bf}

SN~2005bf was an especially interesting example of a Type Ib/c
supernova. It had a double peaked light curve with a first
maximum about 20 days after the explosion and a second, brighter
maximum about 20 days later. In addition, it resembled a helium-poor 
Type Ic in early data, but later developed distinct helium-rich 
Type Ib features (Tominaga et al. 2005; Folatelli et al. 2006;
Parrent et al. 2007).  Spectropolarimetry of this event thus gave 
a chance to look for common aspects of core collapse, clues to 
the very nature of the Type Ib/c spectral classification, and 
evidence concerning the special nature of SN~2005bf itself.

Maund et al. (2007b) presented spectropolarimetry SN~2005bf 
from May 1, 2005, 34 days after the explosion, 18 days after 
the first peak in the light curve, and 6 days before the second 
peak when the supernovae was in the Type Ib phase. Although the 
ISP may have been substantial, intrinsic continuum polarization 
is still consistent with global asymmetry of order 10\%. The 
data tend to fall along a single locus, suggesting a principal axis 
and jet-like flow, but also to show considerable scatter around 
that dominant axis. SN~2005bf showed a distinct loop in the 
He I $\lambda$5876 line (Figure 8) 
adding to the evidence for the ubiquity of such features, perhaps
especially in helium, and further suggesting a departure 
from axisymmetry. Polarization as high as 4\% was observed in the 
absorption components of Ca II H\&K and the Ca II IR triplet, 
rivalling that in SN~1997X. The calcium and helium showed 
different polarization distributions in the Q/U plane. 
Iron lines and the prominant O I $\lambda$7774 line were not
polarized (see SN~2002ap for a contrasting case, \S\ref{grb}). 
SN~2005bf was SP Type D1 overall and SP Type L in the helium line.

The spectropolarimetry helped to construct a picture of the
nature of SN~2005bf that differed from those already
in the literature.
The basic premise of Maund et al. was that the progenitor of
SN~2005bf comprised a C/O core surrounded by a helium mantle;
that is, that the progenitor was structurally that of a Type Ib.  
The lack of Fe polarization suggested that a nickel jet 
had not penetrated to the surface. The lack of O I polarization
suggested that the photosphere had not receded into the C/O-rich core. 
This helium star was presumed to have its own axis of symmetry, 
probably a rotation axis. To that basic picture, Maund et al. added 
the ``tilted-jet." A possible rationale for this assumption is 
given by core-collapse models in which the spin axis of the 
proto-neutron star is de-coupled from the spin axis of the progenitor 
star (Blondin \& Mezzacappa 2007).  In the ``tilted-jet"  model
the continuum polarization arises in the aspherical photosphere. 
The helium mantle contains a solar abundance of calcium. The 
calcium thus resides primarily in the photosphere and reflects
in part the geometry of the photosphere. Some of the calcium is 
also subject to the asymmetric excitation due to the buried 
nickel-rich jet that has penetrated the C/O core but not the 
helium mantle. This gives the calcium a different net polarization 
angle compared to the photosphere. Unlike the calcium, the helium 
is only observed due to excitation of the nickel jet. Since it has 
no photospheric component, it shows yet again another orientation, 
that of the jet. 

Maund et al. suggested that SN~2005bf resembled a Type Ic early 
on because the photosphere had not yet receded to where the
excited helium could be seen. At a later phase, the
excited helium could be seen and the supernova resembled
a Type Ib. The notion that helium might appear in some
events and not others and at different times in different
events has long been discussed (Swartz et al. 1993).
SN~2005bf showed again that spectropolarimetry is a valuable 
probe of the similarities and differences between Type Ib and Type Ic. 

While the tilted jet  model was designed to qualitatively
account for the characteristics of SN~2005bf, its basic features --
an asymmetric core that might vary in mass and composition and
a tilted jet that carries asymmetric excitation to various
depths and at various angles -- may help to illuminate other
core collapse events.  SN~2008D, being observed as this review 
was completed, may be another example of a SN~2005bf-like event. 

\subsection{Polarization of High-Velocity SN~Ic and GRB-related Events}
\label{grb}

As remarked in the Introduction, the gamma-ray burst revolution 
(Woosley \& Bloom 2006) unfolded in close parallel to the developments 
in the polarization of supernovae that we summarize here.  There is 
a strong suggestion that since gamma-ray bursts are the result of jets, 
spectropolarimetry may also shed light on the supernova/gamma-ray 
burst connection (Wang \& Wheeler 1998). There is a concern that 
asymmetric explosions 
could mimic some of the effects of ``hypernova" or ``high-velocity" 
activity that have been interpreted as high energy with spherical 
models (\ho\, Wheeler \& Wang 1999).  Others have argued that large 
ejecta energy and nickel mass are needed even if the explosion were 
asymmetric (Maeda et al. 2002). 

The most famous example of a high-velocity Type Ic, SN 1998bw, 
was apparently associated with the gamma-ray burst of April 25, 
1998. Much has been written of this event (van Paradijs et al. 2000),
a spectacularly bright supernovae and an interestingly dim gamma-ray 
burst.  We will not attempt to summarize that literature here. 
Spectropolarimetry of SN~1998bw was presented by Patat et al. 
(2001) that suggested a moderate intrinsic polarization
that might have increased to the red in the earlier data. 
Few doubt that SN~1998bw was asymmetric. The issue, unresolved, 
remains what affect asymmetries had on the spectral and photometric 
properties. 

The Type Ic SN~2002ap represented an especially interesting case. 
This event showed high velocities, but neither a strong relativistic 
radio source, nor excessive brightness. This event convinced many 
people that the proper nomenclature for this subset of hydrogen- and
helium-deficient supernovae should be ``high-velocity Type Ic,"
rather than ``hypernovae." Spectropolarimetry was obtained by 
Kawabata et al (2002), Leonard et al. (2002a) and Wang et al. 
(2003b) covering about 6 days prior to maximum light to 30 days
after maximum. The data were especially interesting and complex, 
showing a shift in the polarization angle of the continuum and 
different orientation of position angles for oxygen lines, 
calcium lines, iron lines, and the continuum. This is strong 
evidence for non-axisymmetry. SN~2002ap was SP Type D0
in the continuum and SP Type L in strong lines. 

The polarized Ca II IR triplet feature appeared later than that 
of the O I line (Kawabata et al. 2002; Wang et al. 2003b). 
This behavior is expected if the calcium is primarily resident 
in the photosphere, but is excited to display an asymmetric geometry 
only as the ejecta thin out and the outer regions are exposed to the
asymmetric excitation of an underlying nickel jet. Wang et al. (2003b)
proposed that the jet was still buried in the oxygen mantle
in SN~2002ap. A ``buried" jet model might then account for the 
observations of SN~2002ap with the calcium partaking partly of the 
photospheric geometry and partly of the asymmetric excitation, as 
also proposed for SN~2005bf (\S\ref{sn2005bf}). 

SN~2003dh and SN~2006aj were especially interesting as they 
corresponded to the counterparts of gamma-ray burst GRB030329
and the X-ray flash XRF060218, respectively. 
[MARGIN COMMENT X-ray flash events are presumed to be related to 
gamma-ray bursts, but they show softer radiation outbursts.]  
A possible polarization 
of SN~2003dh was reported by Kawabata et al. (2003) at 7000--8000 \AA\ 
(a region contaminated by sky lines) with a position angle of 
$\sim$130\degree, nearly orthogonal to that of the purported 
ISP (Kawabata et al.; their Figure 1d). Since the line of sight was 
closely along the jet of the GRB, presumably a symmetry axis, any 
polarization would represent non-axisymmetry. 

Maund et al. (2007c) reported a single epoch of spectropolarimetry
for SN~2006aj obtained about 10 days after the X-ray flash and hence near 
the estimated maximum of the light curve of the supernova. 
The polarization is nearly 4\% in the blue, making SN~2006aj another 
rival, like SN~1995bf, to SN~1997X (\S \ref{sn1997X}). This raises 
the issue as to whether SN~1997X were an XRF.  Maund et al. note 
that their polarization spectrum is consistent with the photometric 
data of Gorosable et al. (2006) and with the spectropolarimetry of 
Kawabata et al. (2003) of SN~2003dh, in particular the high polarization 
of oxygen and calcium. Once again the event is likely to have 
been observed down the jet axis, and so the polarization is
presumed to be due to a breakdown in axisymmetry.  SN~2006aj was
SP Type L.  

\subsection{Implications of Polarization for Core Collapse}

The fact that the asymmetry of many core-collapse supernovae 
has a tendency to align in one direction provides an important clue 
to the engine of the explosion.  To do what we see, it seems likely 
that the mechanism that drives core-collapse supernovae must produce 
energy and momentum aspherically from the start, then hold that 
special orientation long enough for its imprint to be permanently 
frozen into the expanding matter. Appropriate outflows might be 
caused by magnetohydrodynamic jets, by accretion flow around the 
central neutron star, by asymmetric neutrino emission, by 
magnetoacoustic flux or by some combination of those mechanisms.  
Another alternative, perhaps intimately related, is that material
could be ejected in clumps that block the photosphere in different
ways in different lines. It may be that jet-like flows induce
clumping so that these effects go together.

The light we see from a supernova comes substantially from the decay 
of short-lived radioactive elements, $^{56}$Ni, $^{56}$Co and later
$^{44}$Ti in the debris.  If this material is ejected in an axial  
fashion, then the overall debris shell could be nearly spherical, 
while the asymmetric source of illumination leads to a net polarization. 
This mechanism accounts for the polarization in the early phases of 
Type IIP supernovae.

By injecting jets of mass and energy up and down along a common 
axis, typical aspherical configurations emerge.  
Bow shocks form at the heads of the jets as they 
plow through the core, and a significant portion of the star's 
matter bursts through the core along the jet axis.  The bow shocks 
also drive ``transverse" shocks and associated flow sideways 
through the star. These shocks proceed away from the axis, 
converge toward the star's equator, and collide in the equatorial 
plane. From there, matter is compressed and ejected in an equatorial 
torus perpendicular to the jets. Models have shown that 
sufficiently energetic jets can both cause the explosion of the
supernova and imprint asymmetries (Khokhlov et al. 1999).  Whether 
jets of some sort can alone explain the explosion energetics or 
whether jet-like flow  merely supplements the standard neutrino-driven 
explosion or some other process remains to be seen.  If the jets up and 
down the symmetry axes are somewhat unequal, they might also account 
for the runaway velocities of pulsars and more complex flow patterns 
in the ejecta.  There is a need to understand what happens to the 
dynamics and shape of ejecta if unequal jets break the mirror symmmetry.

\subsection{Summary of Core Collapse Spectropolarimetry}

Every major spectral type of core-collapse supernova has
now been sampled with spectropolarimetry. They are all polarized
and hence substantially asymmetric. This is a general property
of core-collapse, not the peculiarity of single events. While
each individual supernova has its own properties, basic themes 
emerge. The fundamental cause of the asymmetry is deep in the ejecta. 
It is a generic property of core collapse.  The asymmetry is 
characterized by a dominant polarization angle, the most straightfoward 
explanation of which is directed flow, a jet. Atop this basic
structure there are significant, composition-dependent structures
that signal generic, large-scale departures from axisymmetry.
Any physical model of core collapse must address these realities. 

\section{Thermonuclear Supernovae}

The continuum polarization of Type Ia supernovae tends to be
smaller than that for core collapse, but the line polarization
can be substantial. The latter is only clearly observed for
observations prior to maximum light, so the requisite observations 
are challenging to acquire. Although the sample is small, a broad 
range of behavior of Type Ia supernovae has been observed. 
All show substantial asymmetry.

\subsection{``Normal" Type Ia}

SN 1996X was the first Type Ia supernova with  spectropolarimetry 
well prior to optical maximum and the first that revealed a polarized 
component intrinsic to the supernova.  Wang, Wheeler \& H\"oflich (1997) 
presented broadband and spectropolarimetry of SN 1996X about 1 week 
before and 4 weeks after optical maximum. The Stokes parameters derived 
from the broadband polarimetry were consistent with zero polarization. 
The spectropolarimetry showed broad spectral features that were 
intrinsic to the supernova atmosphere with a rather low polarization 
$\sim 0.3 \%$ that were identifiable only after careful theoretical 
modeling.  The spectral features in the polarized spectrum of SN~1996X 
were richly structured and showed no strong correlation with the 
features in the total flux spectrum. Wang et al.  noted that because 
the polarization spectrum is formed at the composition-dependent, and 
hence wavelength-dependent, last-scattering surface the polarized flux 
should show a higher degree of wavelength dependence than the 
multiply-scattered, blended total flux spectrum. Model calculations 
showed that the polarization could be produced by electron scattering 
and scattering from blended lines.  The models suggested that the 
polarization was formed at the boundary between the partially burned 
silicon layers and the inner, iron-peak layers.  No other Type Ia with 
early spectropolarimetry has shown a polarized spectrum quite like SN~1996X. 
The SP Type of SN~1996X is N1.

Leonard et al. (2005) presented spectropolarimetry of 
SN~1997dt about 21 days after maximum light.  This object did 
display distinct line polarization in Fe II and Si II, so some of 
the polarization is intrinsic, but the amount remains uncertain.
As we will see below, the polarization of Type Ia supernovae
at this later phase tends to be quite low.

\subsubsection{SN~2001el: High-Velocity Ca II IR Triplet}
\label{hiVCa}

The first detailed, high quality, time-sampled spectropolarimetry 
of a ``core normal" Type Ia (Branch et al. 2006) was obtained
for SN 2001el (Wang et al. 2003a). A sample of these data is given 
in Figure~10. 
The degree of polarization in the continuum and across spectral 
lines decreased sharply 10 days after optical maximum, and became 
undetectable at about 19 days after maximum.
Prior to optical maximum, the linear polarization of the continuum  
was $\sim 0.2 - 0.3 \%$ with a constant position angle, showing that 
SN~2001el had a well-defined axis of symmetry. Significant 
deviations orthogonal  to the dominant axis were seen suggesting 
that the continuum was SP Type D1. The Si II $\lambda$6355 line showed 
a broad loop on the Q/U diagram and is thus of SP Type L across that 
wavelength range. The polarization was nearly undetectable a week 
after optical maximum.  These data showed that the temporal 
behavior of Type Ia was just the reverse of core-collapse events. 
It is the outer layers of Type Ia that are especially asymmetric, 
not the inner layers.  

SN~2001el also gained fame because it was the first Type Ia
in which the strong high-velocity, 20,000 - 26,000 \kms, 
component of the Ca II IR triplet was observed, a feature 
separated from the photospheric Ca feature. This high-velocity 
feature was identified by Hatano et al. (1999) 
for SN~1994D, but it was much weaker there. The high-velocity Ca II  has 
proven nearly ubiquitous in subsequent studies (Mazzali et al. 2005). 
This high-velocity component showed much higher polarization than the 
continuum, $\sim$ 0.7\% versus 0.2 - 0.3\%, respectively, and a 
different position angle. Because the position angle varies across the 
line, the high-velocity Ca II showed a distinct loop on the Q/U plane
making this feature a dramatic example of SP Type L (see Figure 10).
The polarization spectrum of SN~2003du obtained by 
Leonard et al. (2005) 18 days after maximum was virtually 
identical to that of SN~2001el at a comparable phase,
evidence that this behavior is not isolated to a single
rare event.  Kasen et al. (2003) showed that the loop in the 
high-velocity Ca could be accounted for by a clumpy shell or a 
torus with sufficiently high optical depth (see Figure 3). 

Spectropolarimetry of SN~2004S (Chornock \& Filippenko 2006) 
showed that the high velocity component of Ca II IR was not 
highly polarized although the corresponding position angle was 
rotated with respect to that of the continuum. The relatively 
late phase (9 days after maximum light) lends a cautionary note 
as to whether the low polarization is significant. The behavior
of SN~2004S is consistent with that of SN~2001el and shows
that it is important to acquire early observations of Type Ia supernovae.

\subsubsection{High-Velocity Photospheric Lines}

Another important event was SN~2004dt for which Wang et al. (2006)
obtained data about 7 days before maximum and Leonard et al. (2005)
did so about 4 days after maximum. Although nominally a normal 
Type Ia in terms of its light curve, SN~2004dt falls in a subclass 
of events that show especially high photospheric velocities 
(as opposed to the high-velocity, separated Ca II IR
lines of SN~2001el). This category was discussed by Benetti et al. 
(2004, 2005) as the High Velocity Gradient (HVG) class and by 
Branch et al. (2006) as the broad-line class. 

SN~2004dt also had a distinct polarization spectrum. The variation 
of the polarization across some Si II lines approached 2\%, putting 
SN~2004dt among the most highly polarized SN~Ia.  Unlike SN~2001el, 
for SN~2004dt the polarization of the Ca II IR features were prominent, 
but the polarization levels of the lines of Si II were the strongest. 

Leonard et al. (2005) also present data on SN~2002bf at about the 
same post-maximum time as for their data on SN~2004dt.  Ca II is 
prominent with a polarization of $\sim$ 2\%, with the Si II rather 
modest in the dominant axis.  The Si II line shows distinctly in 
the orthogonal projection, as, perhaps, does a hint of O I. The 
data on SN~2002bf are thus tantalizingly different than those of 
SN~2004dt. Other members of the HVG subclass with spectropolarimetry 
that show strong polarization of the Si II are SN~1997bp and SN~2002bo
(Wang et al. 2006). 

As illustrated in Figure~11, an especially distinct feature of 
SN~2004dt was that, in contrast to the strong polarization of Si II
(upper left panel), the strong line of O I at 7774\AA\ 
showed little polarization signature (lower right panel) 
along the dominant axis defined by the continuum polarization and 
by the strong Si II $\lambda$6355 line.  Despite the large 
evolution in the flux spectrum of SN~2004dt between the premaximum 
observations of Wang et al. (2006) and the postmaximum observations of 
Leonard et al. (2005) 11 days later, there was relatively
little evolution of the polarized spectrum; the Si II, and Ca II IR 
triplet remain significantly polarized and the O I is not.

The SP Types across different spectral lines vary for SN~2004dt, 
as shown in Figure~12.
Thanks to the high quality data from the VLT of ESO, the observational 
errors of SN~2004dt could be determined well enough to enable 
quantitative studies of the Q/U diagram of SN~2004dt. The SI II 
lines and the Ca II IR triplet can be classified as SP Type D1 showing 
distinct dominant axes but with significant dispersion orthogonal
to the dominant axis. The Mg II line is classified as SP Type D0 where 
the Q/U diagram can be well fitted by a straight line.  The O I line 
shows no dominant axis, but with significant dispersion that is too 
large to be explained by observational errors, and is thus classified 
as SP Type N1.  Note that no significant looping is found for the lines. 
This is consistent with a low continuum polarization or only a moderate 
departure from spherical symmetry of the ejecta. The asymmetry we 
see in SN~2004dt is perhaps due to lumpy chemical structure with 
the density of the ejecta more or less spherical, although other 
possibilities could be considered (Livne 1993; Livne, Asida \& \ho 2005). 

The geometrical distributions of different chemical species in 
SN~2004dt are far from spherical. The degree of polarization in 
SN~2004dt points to a silicon layer with substantial departure from 
spherical symmetry. A geometry that would account for the observations 
is one in which the distribution of oxygen is on average
essentially spherically symmetric, but with protrusions of 
intermediate-mass elements within the oxygen-rich region. 
The central regions of SN~2004dt are essentially spherical and 
consist of fully burned iron- group elements.  The 
distribution of magnesium, of SP Type D0, shows a well-defined 
symmetry axis and rather smooth geometry (see Figure~1) when 
compared to the distribution of Si and Ca. The prominent dominant
axis of SN~2004dt in Si, Ca, and Mg may be the result of an 
off-centered ignition (Livne 1999; Gamezo et al. 2004; Livne, Asida 
\& \ho\ 2005; R\"opke, Woosley \& Hillebrandt 2007; Plewa, Calder \& 
Lamb 2004; Plewa 2007), or somehow related to the progenitor system.

Leonard et al. (2005) point to the simulations of Travaglio et al. (2004)
that had significant unburned C and O as a basis for 
interpretation.  As stressed by Wang et al.
(2006), the O has the same velocity profile as the Si in the 
total flux spectrum, a basic fact that is inconsistent with
the model of Travaglio et al., where the silicon remains interior
to the unburned oxygen. Wang et al. discuss the inconsistencies
of their observations with pure deflagration models that
leave substantial unburned C and O.

Another well-studied event of the HVG kind 
is SN~2006X. This event was especially marked as the 
source of the first distinct evidence for a circumstellar medium 
near a normal Type Ia by Patat et al. (2007). 
Patat et al. (2008) report 8 epochs of spectropolarimetry of 
SN~2006X prior to maximum light and one epoch about 40 days later. 
The polarization of SN~2006X uncorrected for ISP shows a 
linear decline from about 8\% at 4000 \AA\ to about 2\% at 8000 \AA. 
This cannot easily be fit with a Serkowski law, implying that the 
dust around SN~2006X is not of the same constitution as the average 
in our Galaxy.  In the earliest data, 10 days before maximum light, 
the Ca II IR line is strongly polarized, $\sim$1.5\%, and the 
Si II less so, $\sim$0.5\%.  As the peak is approached, the Ca gets 
less polarized, the Si more  so. Especially interesting is that the 
Ca polarization seems to be stronger at +39 days than it had become 
at day -1.  This may be due to seeing the inner, clumpy, calcium-rich
layers. No other SN~Ia has been observed at this epoch with 
this level of S/N ratio, so it is not known if this is a common 
property, but it may be. More late-time observations are called for.
Overall, SN~2006X seems to be intermediate in its polarization
properties between SN~2001el and SN~2004dt. Like SN~2004dt, 
SN~2006X shows little or no polarization of the O I line,
less than 0.3\%. The continuum shows a reasonably well-defined
dominant axis in the Q/U plane, but with significant,
and probably real departures. SN~2006X suffers strong interstellar 
reddening and it is likely that the dust responsible for the 
reddening is distributed close to the supernova itself (Wang et al. 2007). 

It is interesting to point out that all the high-velocity Type Ia show 
strong polarization, and many of them show strong dust extinction. 
The interpretation of the polarization data for these supernovae 
may be complicated by the presence of dust particles around the 
supernova (Wang, 2005). 

\subsection{Polarization of Subluminous Type Ia}

Another class of Type Ia that can be studied with spectropolarimetry 
are those in the low-luminosity category, the prototype of which is 
SN~1991bg.  SN~1999by, observed by Howell et al. (2001), is a nice 
example of a sub-luminous Type Ia. The continuum polarization was 
substantial (0.3 - 0.8\%) and rose toward the red as predicted by 
models in which the line scattering depolarization is less in the red. 
The data show a distinct locus in the Q/U plane indicating a 
favored orientation of the geometry, and hence an SP Type D0.
It is not yet clear whether this is a distinguishing characteristic 
of sub-luminous Type Ia. 
Howell et al. speculate that the relatively high, well-oriented
polarization might be the signature of rapid rotation or
binary merger that was a characteristic of sub-luminous events.
Howell et al. note that for core normal Type Ia, the
photosphere should be near the inner iron-rich layers near
maximum light and that these iron lines will affect the spectrum.
For a sub-luminous event like SN~1999by near maximum, the 
photosphere is expected to be still in the silicon layers that 
are more dominated by continuum scattering. 

\subsection{The Extremes of Type Ia}

\subsubsection{The Hybrid Nature of SN 2002ic}
\label{SN2002ic}

SN~2002ic showed distinct evidence for narrow-line hydrogen emission 
similar to Type IIn and indicating strong circumstellar interaction, 
but with an underlying spectrum that was very similar to a Type Ia  
(Hamuy et al. 2003).  SN~2002ic is thus a very intersting
sub-class of Type Ia that have managed, by some circumstance,
to explode in a region still rich in hydrogen (see the discussion
of SN~2006X above, and the discussion in Gerardy et al. (2004) as
to whether the high-velocity Ca II feature represents a swept
up hydrogen shell).  Hamuy et al. conclude that the progenitor system 
contained a massive asymptotic-giant-branch star that lost several solar 
masses of hydrogen-rich gas before the supernova explosion.  

Wang et al. (2004) obtained spectropolarimetry of SN~2002ic nearly 
a year after the explosion.  This is the latest phase of any 
supernova for which a positive detection of polarization
has been made. At this phase, the supernova had become fainter overall, 
but the H$\alpha$ emission had brightened and broadened dramatically 
compared to earlier observations.  The spectropolarimetry 
showed spectropolarization properties typical of those of Type IIn 
supernova, and indicated that the hydrogen-rich matter was highly 
aspherically distributed.  Wang et al. argued that the narrow peak 
and broad wings of the \Ha\ line and the increase in strength of \Ha\ 
with time may be produced by electron scattering in a shocked-heated 
nebular torus.  Chugai \& Chevalier (2007) modeled the observed 
polarization in more detail and concluded that a moderate degree 
of asymmetry with the major axis about 40-50\% longer than the 
minor axis is sufficient to explain the polarization data. 
We remark here that the amount of asymmetry deduced by 
Chugai \& Chevalier (2007) represents the minimum level of 
asphericity required to explain the data, and the level is still 
very significant.  The observations are consistent with about a solar 
mass of clumpy material extending to $\sim 3\times10^{17}$ cm. 
These observations suggest that the supernova exploded inside 
a dense, clumpy, disk-like circumstellar environment very 
reminiscent of a proto-planetary nebula.

\subsubsection{SN 2002cx and SN 2005hk: Weird Type Ia?}

SN~2002cx and SN~2005hk were labeled as a ``very peculiar" Type Ia 
showing low peak luminosity, slow decline, high ionization near peak 
and an unusually low expansion velocity of only about 6,000 \kms (Li et al.
2003; Jha et al. 2006; Chornock et al. 2006; Phillips et al. 2007).  
The low peak luminosity and slow expansion have been interpreted as 
possible evidence for an explosion by pure deflagration. Such an 
explosion is predicted by models to burn less of the white dwarf and 
to produce less nickel to power the light curve, but in published 
models this should lead to a significant amount of unburned carbon 
and oxygen. There is little evidence for that unburned material. 

Chornock et al. (2006) presented spectropolarimetry about 4 days 
before maximum of SN~2005hk. They found that the continuum was 
polarized about 0.4\% in the red with a single axis of 
symmetry dominating the geometry, but with considerable scatter. 
They also identified a weak modulation of the polarization of the 
Fe III line at 5129 \AA. They argued that the  fairly large 
continuum polarization and lack of strong line features are not in
keeping with the predictions of the model of Kasen et al. (2004)
for a Type Ia viewed ``down the hole" carved by a binary
companion in the supernova ejecta (Marietta, Burrows \&
Fryxell 2000). SN~2002cx, SN2005hk, and their kin still defy 
consistent explanation.

\subsection{The Polarization of the Si II 635.5 nm Lines}

The Si II 6355 \AA\ line of Type Ia supernovae is one of the 
strongest lines in the optical range. Wang, Baade, \& Patat (2007) 
reported polarization data of 17 Type Ia and compared the degree 
of polarization across the Si II 6355 \AA\ line to the light curve 
properties of the supernovae. A general trend was identified that 
shows that the supernovae with faster light curve decline rate 
past optical maximum tend to be more highly polarized. The 
correlation shows a large dispersion due to observational noise, 
but the correlation appears to be significant. The linear fit 
gives $P_{SiII}\ = \ 0.48(03) + 1.33(15)(\Delta m_{15} - 1.1)$, 
where $\Delta m_{15}$ is the magnitude decline from peak at 15 days 
after $B$-band maximum.  The supernova showing the highest 
polarization across the Si II 635.5 nm line is SN~2004dt, which does 
not follow the linear fit. 

Since the continuum polarization is low for all these supernovae, 
it is clear that the overall structure of Type Ia supernovae must 
be rather spherical. This is a nice check of using Type Ia as standard 
candles for cosmological studies. 

If we accept $\Delta m_{15}$ as a measure of the brightness of 
Type Ia supernovae (Phillips, 1993), and that brighter Type Ia have 
a larger amount of $^{56}Ni$, we can conclude that supernovae with 
lower polarization have gone through more complete nuclear burning. 
This may be reasonable since  more complete burning tends to destroy 
more inhomogeneities and thus makes more spherical supernovae.

\subsection{The Implications of Polarization for Thermonuclear Explosions}

The first polarization models of Type Ia supernovae were presented 
in Wang et al.  (1997).  These were based on ellipsoidal variations 
of one-dimensional thermonuclear combustion models.  
Kasen et al. (2003) presented models of SN~2001el that explored 
the polarization properties, especially of the high-velocity Ca II IR
triplet that was highly polarized with a different polarization angle 
than the rest of the spectrum.  Kasen et al. investigated the 
non-axisymmetric geometries necessary to produce ``loops" in the Q/U 
plane. A clumped shell model did an especially good job of accounting 
for the features of the loop defined by the high-velocity Ca II IR line
(see Figure 2 for an example of this sort of calculation), but such
models remain phenomenological. A deeper understanding remains 
elusive. 

Wang et al. (2006) discussed in the context of various models  
the challenging observations of SN~2004dt that showed polarized 
Mg, S, and Si, but nearly unpolarized O. This is a puzzle because models
predict that these elements should be intermixed in the outer 
layers of the explosion and hence might be expected to share 
geometry.  One important class of models are deflagration models
in which the burning is driven only by subsonic turbulent flames
that produce irregular structure, but do not burn the entire 
C/O white dwarf (Reinecke, Hillebrandt, \& Niemeyer (2002); 
Gamezo et al. 2003; R\"opke et al. 2007). Wang et al. concluded 
that while turbulent deflagration models might produce the observed 
clumpy composition structure, they have difficulty producing the 
kinematics of the ejecta, especially the high-velocity yet
asymmetric silicon. Another important class of models are the
delayed-detonation models (Khokhlov 1991) in which the burning 
starts as a turbulent, subsonic flame, but there is a presumed 
transition to a supersonic, shock-driven detonation. These models 
provide good fits to light curve and spectral data and account for
the energetics.  Delayed-detonation models burn the outer layers 
to intermediate mass elements, but are unlikely to generate a 
turbulent silicon layer as apparently observed in SN~2004dt 
(Boisseau, et al. 1996; Gamezo et al. 1999). Off-center 
delayed-detonation models (Livne 1999; Gamezo et al. 2004;
Livne, Asida \& \ho\ 2005; R\"opke, Woosley \& Hillebrandt 2007) 
can yield off-center nickel and off-center, clumpy silicon that 
may contribute to the induction of polarization (Chugai 1992; 
H\"oflich 1995a), but there is, as yet, no completely satisfactory 
model for the polarization of SN~2004dt. 

A related possibility is the gravitationally-confined detonation 
model (Plewa, Calder \& Lamb 2004) or detonating failed deflagration 
model (Plewa 2007) in which a plume of burned matter floats to the 
surface and across the surface of the star to converge at the point 
opposite from where the plume emerges. The resulting rapid compression
of the unburned material is proposed to trigger a detonation.  This 
type of model has been explored in some detail by Kasen \& Plewa (2007). 
They generate synthetic broadband optical light curves, near-infrared 
light curves, color evolution curves, full spectral time series, 
and spectropolarization of the model, as seen from various viewing 
angles. They compare the statistical properties of the model to 
observations and comment on orientation effects that may contribute 
to photometric and spectroscopic diversity. Kasen \& Plewa remark
that their model, although based on a particular physical
scenario, may be characteristic of a general class of asymmetric
models. They made no attempt to compare to a particular supernova,
but their total flux spectra do not give a significant O I line and 
hence their model does not correspond directly to SN~2004dt.   

\subsection{Summary of Type Ia Spectropolarimetry}

Type Ia supernove are more polarized in the outer layers than
the inner. This is an important clue to the nature of their
thermonuclear burning.  The continuum polarization is low showing that
the explosion is nearly spherical, but the line polarization
can be very strong.  The line polarization may correlate with
the velocity measured from absorption line minimum (Leonard
et al. 2005). Some Type Ia show significant orientation with a
dominant axis, but it is not yet clear whether this correlates
with other properties, for instance the light curve decline
rate. The future holds great promise to use spectropolarimetry
of Type Ia to explore the asymmetries of the explosion, 
the nature of the progentor system, which should be 
intrinsically asymmetric, and the nature and geometry of
the circumstellar medium. 

\section{Conclusions}

Spectropolarimetry of supernovae is still a young field, but
a number of basic conclusions can be reached that have
important implications for the future of supernova research:

$\bullet$ Spectropolarimetry has now been obtained for every major 
spectral type of supernova: Type IIP, Type IIn, Type IIb,
Type Ib, Type Ic, and Type Ia of various luminosity classes
and peculiarities (Table 1). They are all polarized and hence 
aspherical in some significant way.

$\bullet$ Presenting the data in the Q/U plane as a function of time,
wavelength and especially across indvidual spectral lines
is a powerful way to facilitate analysis of the data.

$\bullet$ Core-collapse supernovae reveal larger asymmetry as observations 
are made deeper in the ejecta, either as the ejecta thins with 
expansion or the progenitors have less massive hydrogen envelopes. 
This is firm evidence that the explosion mechanism itself is 
strongly asymmetric.

$\bullet$ Type Ia supernovae display modest continuum polarization
but very strong line polarization prior to maximum light.
The polarization declines after maximum. This is firm
evidence that the outer portions of the ejecta are
especially significantly aspherical.

$\bullet$ Core-collapse supernovae routinely show evidence for 
strong alignment of the ejecta in single well-defined
directions, suggestive of a jet-like flow.

$\bullet$ Core-collapse supernovae often show a rotation of the
position angle with time of 30 -- 40\degree\ that is
suggestive of a jet of material emerging at an angle
with respect to the rotational axis of inner layers.

$\bullet$ Core-collapse supernovae frequently display loops
in the Q/U plane that indicate significant and
systematic departures from the predominant axisymmetric 
structure. 

$\bullet$ SN~1987A shows all of the features displayed by core-collapse
supernovae: strongly directed flow, but also significant
evidence for non-axisymmetric structure. SN~1987A is
worthy of in-depth reconsideration in the current context.

$\bullet$ Spectropolarimetry is an especially sensitive tool to
study the presence and asymmetry of helium that is
excited by non-thermal processes associated with
radioactive decay. The helium, in turn, is an important
clue to the nature and extent of mass loss from the
progenitor, to the relations between Type IIb, Type Ib,
and Type Ic supernovae, and to the asymmetric distribution 
of the radioactive elements.

$\bullet$ Spectropolarimetry of supernovae can reveal information
about the properties of the interstellar dust in the host galaxy. 
Current data suggest that this dust frequently has properties 
different than dust in the Galaxy.  

$\bullet$ Large scale asymmetries in the outermost, high-velocity
portions of the ejecta of Type Ia supernovae are not
consistent with pure deflagration models, but also
not simply explained by delayed-detonation models. 
This remains a key challenge to models of thermonuclear
explosions.

H\"oflich et al. (1996) argue that to some order the flux spectra 
of aspherical models can be reproduced by appropriate adjustment 
of the parameters of spherical models, for instance the radius of 
the photosphere, the density gradient, or perhaps some 
``microturbulence," so that it is difficult to deduce the geometry 
from the early time flux spectra of supernovae. Certainly flux spectra 
are mute to important aspects such as different orientations for 
different chemical species and ``loops" in the Q/U plane. 
Nebular spectra may be more useful indicators of asymmetry since 
line profiles can reveal velocity asymmetries and small scale 
clumping structure, but again, there is no information on the 
relative orientatation of different chemical species. 

The totality of the evidence derived from spectropolarimetry of 
core-collapse supernovae is that the primary driver of asphericity
is deep in the explosion process and that the explosion process 
often ``aims" at a certain direction in space. Something resembling
jet-like flow is strongly suggested. 

As we have summarized here, recent work has revealed that 
departures from axisymmetry are also ubiquitous in core-collapse 
supernovae. Much more needs to be done to quantify this significant 
clue to the physics and dynamics. Two basic means to induce asphericity 
are from a distortion of the progenitor and a jet-like flow that
can induce asphericity in both the density structure and
the ionization structure. Jet-like flow is likely to induce
significant non-axisymmetric hydrodynamic instabilties: Rayleigh-Taylor,
Kelvin-Helmholz, and Richtmyer-Meshkov (Wheeler, Maund \& Couch 2008).
A non-axisymmetric geometry could also be achieved if the jet-like flow 
were tilted with respect to the progenitor geometry. There are 
physical reasons why the rotation axis of a newly-born neutron star 
and hence, presumably, the jet, could be different than the 
rotational axis of the progenitor.  The jet could also penetrate 
to different depths in different circumstances, giving considerable 
variety to the basic picture.  In addition, clumps of different 
density or composition could cause or contribute to the non-axisymmetry. 
Ultimately there is also a need to understand the origin of such clumps
with the jet being a potential principal source.

It remains true that the characteristics of ``high-velocity"
Type Ic may be the result of orientation effects in a mildly 
inhomogeneous set of progenitors, rather than requiring an excessive 
total energy or luminosity.  In the analysis of asymmetric events with 
spherically symmetric models, it is probably advisable to refer 
to ``isotropic equivalent" energy, luminosity, ejected mass, 
and nickel mass. The discovery that routine core-collapse
supernovae are aspherical and ``jet-like" may, in turn, give 
new insights into more exotic jet-induced events like gamma-ray bursts
and X-ray flashes.  
We desperately need more high quality spectropolarimetry of 
stripped-core events.  This has proved difficult in practice, but 
the few that have been observed have been exceedingly interesting.

Application of SN~II to measurements of the distance scale need to 
be considered with care since even asymmetric luminosity sources 
in the plateau phase can give a small direction dependence to the 
luminosity (Chieffi et al. 2003).

The asymmetries observed in Type Ia may finally yield direct 
observational evidence that they occur in binary systems, as 
long assumed, and clues to the combustion mechanism.  
At this writing, there is no fully acceptable model of
the origin of the composition-dependent aspherical structure
of Type Ia supernovae, certainly not the possible variation
of that structure within the distribution of other observed 
properties of Type Ia. In particular, neither pure deflagration 
nor delayed detonation nor gravitationally-confined detonation
models fully account for the observations. On the contrary,
the spectropolarimetry has offered a new window by which
to characterize the nature of the explosion. As the 
sample grows, we should more deeply understand the nature
of the explosion and how it varies among the normal, luminous, 
and subluminous Type Ia, and among the high-velocity gradient, 
broad-line, low-velocity and other sub-genres of the species,
such as the hydrogen-rich SN~2002ic and the especially ``weird" 
SN~2002cx. The seemingly ubiquitous, substantially polarized
high-velocity Ca II feature is an especially important aspect
to explore with spectropolarimetry.
Understanding these asymmetries may be necessary to properly 
interpret future data on cosmologically distant Type Ia's.

This topic is still in a data-driven phase, but there has been 
much excellent work to lay the ground for understanding (Jeffery 
1989, 1990, 1991b; \ho\ 1991; Chugai 1992, 2001, 2006; Kasen et al. 
2003; Kasen et al. 2006).  Much of the basic understanding of the 
possible behavior of polarized supernovae, especially of the 
continuum, has been elucidated in terms of ellipsoidal models 
(\ho\ 1991; \ho\ 1995a; \ho\ et al. 1996). Prolate models tend 
to require a larger distortion than oblate models to produce the 
same net polarization (H\"oflich et al. 1996). The true geometry 
may, of course, be some combination of oblate and prolate and 
more complex than captured easily by ellipsoidal structures.

Exploring more complex structures requires more general models. 
Monte Carlo radiative transfer models are the most likely to
provide the needed generality of the underlying geometry and
the capacity to handle the associated radiative transfer with
a realistic amount of computer time. Contemporary examples
of this are the HYDRA code written by \ho\ (2005) and the SEDONA 
code written by Kasen (see Kasen et al. 2006) which have been
applied to a number of problems, especially those associated
with Type Ia. These codes treat aspherical geometries, gamma-ray
deposition, and line and continuum transfer.  They can compute models
for the light curves and for the spectral and polarimetric evolution 
as observed from arbitrary viewing angles. 

There are two ways to approach the
analysis of polarization data. One is the solution of the 
``forward" problem for which a dynamical model is proposed or 
simulated and the polarimetric evolution is computed to compare 
with observations. The solution of the ``backward" problem 
consists of proceeding from the observed data to deduce the
geometry of the ejecta. In practice, progress will be made
by a process of iteration between these two techniques.
Codes such as these will play a vital role in this iteration. 
This phase of supernova polarimetry research is just beginning,
but holds great promise as the data base expands.

\section{Acknowledgments}
The authors are grateful to their fellow members of the
VLT spectropolarimetry team for so much of the hard work
and insight that made this review possible: PI Dietrich Baade, 
Peter H\"oflich, Nando Patat, Alejandro Clocchiatti, and 
Justyn Maund. Special thanks go to Peter H\"oflich for his 
modeling work and early support of the Texas spectropolarimetry 
program when it was first getting off the ground. JCW is
also especially grateful to Justyn Maund for extensive
and insightful discussions of supernova polarimetry and 
related topics and for his help with some of the graphics
used here. We also thank Doug Leonard for helpful discussions 
and for providing specially crafted figures representing his 
work and Dan Kasen for comments and a key figure. We are 
indebted to the European Southern Observatory for the generous 
allocation of observing time for our VLT
program and to The Paranal Science Operations staff and the User 
Support Group in Garching who have gone to considerable effort 
implement our demanding program. We recognize that accommodating 
target-of-opportunity observations in an already busy observing 
and work schedule often poses a special extra challenge. This work 
was supported in part by NSF Grant NSF AST 0709181 to LW and 
NASA Grant NNG04GL00G and NSF Grant AST 0707769 to JCW.

\clearpage

\begin{table} \scriptsize
\def~{\hphantom{0}}
\caption{Supernovae with Polarimetric Data}\label{snpol_data}
\begin{tabular}{llccll}
\toprule

Supernova & Host Galaxy & Type  & Photometric (P) & No. of Epochs: epoch\footnote{with respect to optical maximum} & References\\  
& & & Spectroscopic (S) & (SP Type) & \\  
\colrule

1968L   & M83 (NGC 5236)&IIP&P (undetermined)&1&1 \\
1970G   & M101 (NGC 5457)&IIL&P&1&2\\
1972E   & NGC 7723&Ia&P (undetermined)&3&3\\
1975N   &NGC 7723&Ia&P? (undetermined)&2: 0, 34&4\\
1981B   &NGC4536&IA&P (undetermined)&1: 56&5\\
1983G   &NGC 4753&Ia&S? (undetermined)&1: -2&6\\
1983N   &NGC 5236 (M83)&Ib&S&1: 1&6\\
1986G   &NGC5128&Ia&S&2: -9, -8&7\\
1987A   &LMC&IIpec&P, S&many&8\\
1992A   &NGC 1380&Ia&S (undetermined)&1: 12 (N) &9\\
1993J   &M81&IIb&P, S&17 (D1, L): -15, -14, -11 & 10\\
     &   &   &    & +2(x2), +8, +9, +10, +12\\
     &   &   &    & +23, +26\\
     &   &   &    &  plus 6 photometry\\
1994D   &NGC 4526&Ia&P (undetermined)&1: -10&11\\
1994Y   &NGC 5371&IIn&P&1: $>$180 (D0?) &11\\
1994ae &NGC 3370&Ia&P (undetermined)&1: $>$30 (N0) &11\\
1995D   &NGC 2962&Ia&P (undetermined)&2: 14 (N0), 41 (N0)&11\\
1995H   &NGC 3526&II&P&1 $>33$&11\\
1995V   & NGC 1087 &II&P?&n&12\\
1996B   & NGC 4357 &IIb&S&1?& 13\\
1996W  & \ldots &II&S?&n&12\\
1996X   &NGC 5061&Ia&S&1 (N1) &14\\
1996cb & NGC 3510 &IIb&S&n (D1, L) &15\\
1997X & NGC 4691 &Ic&S&2: $\sim$5, $\sim$30 (D1, L1)&15\\
1997Y   & \ldots &Ia&S (undetermined)&n (N0) &12\\
1997bp & NGC 4680 &Ia&S?&n (D0) &12\\

\botrule
\end{tabular}
\end{table}

\setcounter{table}{0}
\begin{table} \scriptsize
\caption{Continued}
\def~{\hphantom{0}}
\begin{tabular}{llccll}
\toprule

Supernova & Host Galaxy & Type  & Photometric (P) & No. of Epochs: epoch\footnote{with respect to optical maximum}  & References\\  
& & & Spectroscopic (S) & (SP Type) & \\  
\colrule

1997bq & NGC 3147 &Ia&S (undetermined)&1 (D1) &12\\
1997br  & \ldots &Ia&S (undetermined)&1 (D1) &12\\
1997dq& NGC 3810 &Ib&S&1 (D1) &16\\
1997ds&MCG -01-57-007&II&S&1 (D1) &17\\
1997dt&NGC 7448&Ia&S&1 (N1) &18\\
1997ef& UGC 4107 &Ib/c (high velocity)&S (undetermined)&2? (N1) &19\\
1997eg&NGC 5012&IIn&S&3: 16, 44, 93 post discovery (D1) &20\\
1997ei&  NGC 3963 &Ic&S&1&12\\
1998A&IC 2627&II&S&1&17\\
1998S&NGC 3877&IIn  &S&3: -14, 10, 41 (D)&21\\
1998T& NGC 3690 &Ib&S&1&16\\
1998bw&-&Ic (pec)&S&2: -6, 10 (N) &22\\
1999by&NGC 2841&Ia (subluminous)&S&3: -2, -1, 0 (D1) &23\\
1999em&NGC 1637&IIP&P, S&7: 7, 10, 40, 49, 73 (D), &24\\
     &   &   &    & 159, 163 (after discovery)\\ 
1999gi&NGC 3184&II&S&1 (D, L) &17\\
2001V&NGC 3987&Ia&S&1 (D1, N) &25\\
2001X&NGC 5921&IIP&S&1 (D) &25  \\
2001bb&??&Ic&S&1(D1) &25\\
2001dh&MCG -6-44-26&II P?&S&2 (D, L) &25\\
2001dm&??&Ia&S&1 (N, D) &25\\
2001du&NGC 1365&IIP&S&1 (N, D) &25\\
2001el&NGC 1448&Ia&S&5: -4 (D1, L), +2 (D1, L) &26\\
     &   &   &    & +11 (D), +20 (N1), +41 (N1)\\
2001ig&NGC 7424&IIb&S&3:+13 (D, L), +31 (D, L), +256 &27\\

\botrule
\end{tabular}
\end{table}


\setcounter{table}{0}
\begin{table} \scriptsize
\caption{Continued}
\def~{\hphantom{0}}
\begin{tabular}{llccll}
\toprule

Supernova & Host Galaxy & Type  & Photometric (P) & No. of Epochs: epoch\footnote{with respect to optical maximum}  & References\\  
& & & Spectroscopic (S) & (SP Type) & \\  
\colrule

2002ap&M74 NGC 628&Ic (HV)&S&S: -6 (, L), -2 (N, L), 0 () &28\\
    &   &   &    & +1(x2) (D, L), +2,  \\
    &   &   &    & +3(x2), +5, +26,  +27, +29\\
2002bo&NGC 3190&Ia&S&2 (L) &25\\
2002el&NGC 6986&Ia&S&1 (D) &25\\
2002fk&NGC 1309&Ia&S&2 (D) &25\\
2002bf&??&Ia&S&1 (N) &18\\
2002ic&??&Ia (pec)&S&5: 221 (D1), 232 (D1) &29\\
    &   &   &    & 253 (D1), 255 (D1), 315 (D1)\\
2003B&NGC 1097&II&S&1 (D) &25\\
2003L&NGC 3506&Ic&S&1 (D) &25\\
2003W&UGC 5234&Ia&S&1 (D) &25\\
2003bu&??&Ic&S&2 (D) &25\\
2003dh&a104450+213117&Ic (HV)&S&2: 15 (N), 16 (N) &30\\
2003du&UGC 9391&Ia&S&1 (D) &18\\
2003ed&NGC 5303&IIb&S&1&38\\
2003ee&??&IIn&S&1 (D1) &25\\
2003eh&??&Ia&S&1(N1) &25\\
2003gd&M74 NGC 628&IIP&S&1 (D) &25\\
2003gf&MCG -04-52-026&Ic&S&1&38\\
2003hh&??&Ia&S&2 (D1) &25\\
2003hv&NGC 1201&Ia&S&1 (D) &25\\
2003hx&NGC 2076&Ia&S&1 (D) &25\\
2003hy&IC 5145&IIn&S&1 (D1) &25\\
2004S&MCG-05-16-021&Ia&S&1 (D) &31\\
2004dj&NGC 2403&II&S&1 (D) &32\\

\botrule
\end{tabular}
\end{table}


\setcounter{table}{0}
\begin{table} \scriptsize
\caption{Continued}
\def~{\hphantom{0}}
\begin{tabular}{llccll}
\toprule

Supernova & Host Galaxy & Type  & Photometric (P) & No. of Epochs: epoch\footnote{with respect to optical maximum}  & References\\
& & & Spectroscopic (S) & (SP Type) & \\
\colrule

2004dk&NGC 6118&Ib&S&1 (D) &25\\
2004dt&NGC 799&Ia&S&2 -7 (D1, L), +4 (D1, L) &33\\
2004br&NGC 4493&Ia (pec)&S&1 (N1) &25\\
2004dk&NGC 6118&Ib&S&1 (D1) &25\\
2004ef&UGC 12158&Ia&S&1 (D1) &25\\
2004eo&NGC 6928&Ia&S&1 (D1) &25\\
2005bf&MCG+00-27-005&Ib/c&S&1: -6 (wrt 2nd max) (D1, L) &34\\
2005cf&MCG -1-39-3&Ia&S&1 (N1) &25\\
2005de&UGC 11097&Ia&S&2 (N1) &25\\
2005df&NGC 1559&Ia&S&2 (D1) &25\\
2005el&NGC 1819&Ia&S&1 (D1) &25\\
2005hk&UGC 272,&Ia (pec)&S&3: -4 (D1), 0 (D1) , +14 (D1) &35\\
2005ke&NGC 1371&Ia&S&2 (D1) &25\\
2006X&NGC 4321&Ia&S&9: -10 (D1, L), -8 (D1, L)&36\\
    &   &   &   & -7 (D1, L), -6 (D1, L), -3 (D1, L)\\
    &   &   &   &  -2 (D1, L), -1 (D1, L), +39 (N1)\\
2006aj&a032140+165203&Ib/c (XRF)&P, S&9: -7 (N1), -6 (N1), -5 (N1) &37\\
    &   &   &   & -4 (N1), 0 (N1), +3 (N1)\\
    &   &   &   & +8 (N1), +29 (N1), 300?? (N0)\\
2006bc&NGC 2397&II&S&1 (D)&25\\
2006be&IC 4582&II&S&1 (D) &25\\
2007af&NGC 5584&Ia&S&2&25\\
2007fb&UGC 12859&Ia&S&4&25\\
2007hj&NGC 7461&Ia&S&3&25\\
2007if&anon&Ia&S&5&25\\
2007it&UGC 10553&II&S&3&25\\
2007le&NGC 7721&Ia&S&3&25\\
2008D&NGC 27720&Ib/c (pec)&S&2 (D1) &25\\

\botrule
\end{tabular}

\end{table}

\clearpage

\begin{table}

\flushleft {\scriptsize References - (1) Wood \& Andrews (1974): Serkowski (1970); (2) Shakhovskoi \& Efimov (1972); (3) Wolstoncraft \& Kemp (1972),  Lee et al. (1972), Shakhovskoi (1976); (4) Shakhovskoi (1976); (5) Shapiro \& Sutherland (1982); (6) McCall et al. (1984), McCall (1985); (7) Hough et al. (1987); (8) Cropper et al. (1988), M\'endez et al. (1988); (9) Spyromilio \& Bailey (1993); (10) Trammel, Hines, \& Wheeler (1993), Doroshenko, Efimov \& Shakhovskoi (1995), Tran et al. (1997) ; (11) Wang et al. (1996); (12) Wheeler (2000); (13) mentioned in Wang et al. (2001); (14) Wang, Wheeler \& H\"oflich (1997); (15) Wang et al. (2001); (16) Leonard, Filippenko, \& Matheson (2000); (17) Leonard \& Filippenko (2001); (18) Leonard et al. (2005); (19) Leonard et al. (2000); Wheeler (2000);
(20) Leonard et al. (2000); (21) Leonard et al. (2000); Wang et al. (2001); (22) Patat et al. (2001); (23) Howell et al. (2001); (24) Leonard et al. (2001); Wang et al. (2001); (25) unpublished; (26) Wang et al. (2003a); (27) Maund et al. (2007c); (28) Kawabata et al. (2002), Leonard et al. (2002), Wang et al. (2003); (29) Wang et al. (2004), Kawabata et al. unpublished; (30) Kawabata et al. (2003); (31) Chornock \& Filippenko (2006); (32) Leonard et al (2006); (33) Leonard et al. (2005), Wang et al. (2006); (34) Maund et al. (2007a); (35) Chornock et al. (2006), Maund et al. (2007d) ; (36) Patat et al. in prep; (37) Gorosabel et al. (2006), Mazzali et al. (2007), Maund et al. (2007c); (38) Leonard \& Filippenko (2005)}\\

\end{table}

\clearpage

\begin{figure}
  \label{QUdominant}
  \centerline{\psfig{figure=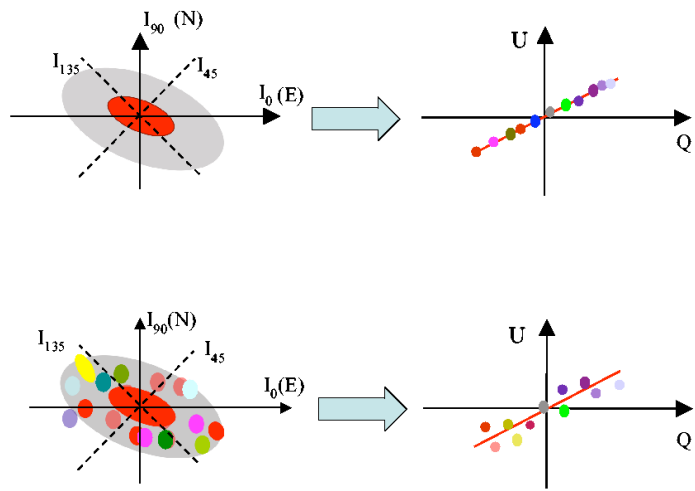,height=0.4\textheight,angle=90}}
\caption{
The top left illustrates a smooth, axisymmetric structure with
the axis tilted at an arbitrary position angle on the sky.
The directions denoted by I represent the measurement of the
flux at the angles needed to construct the Q and U
polarization components (\S \ref{stokes}).  The resulting 
wavelength-dependent polarization amplitude plotted in the Q/U plane 
will follow a straight line, the dominant axis, as illustrated in 
the top right. Dots represent the polarization measured at different 
wavelengths.  The lower left illustrates a case for which the axisymmetry 
is broken into ``clumps" of different composition and optical depth.
Clumps of high-opacity, absorbing material will block
parts of the underlying, photosphere. This will induce
wavelength-dependent geometry. The polarization distribution in 
the Q/U plane will in general no longer be along a straight line, as
illustrated in the lower right. The basic axisymmetric geometry 
may still be evident, as illustrated, but the departure from
axial symmetry caused by the ``clumping" will yield a finite,
and physically significant, distribution along the orthogonal axis.
The Q/U diagrams on the upper and lower right represent 
spectropolarimetry Types D0 and D1, respectively, as defined
in \S \ref{types}.
}
\end{figure}

\clearpage

\begin{figure}[h]
  \label{Kasen_loops}
  \hfill
\begin{minipage}[t]{.45\textwidth}
      \begin{center}
      \psfig{figure=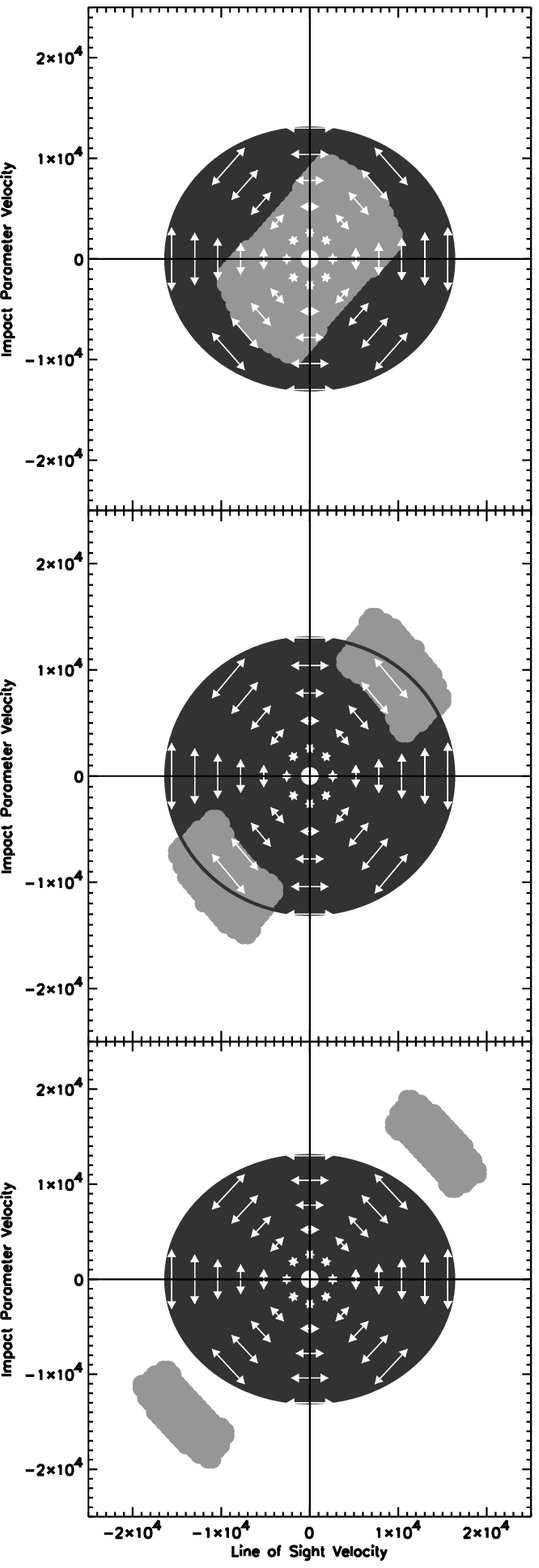,height=0.4\textheight}
      \end{center}
  \end{minipage}
  \hfill
  \begin{minipage}[t]{.45\textwidth}
      \begin{center}
      \psfig{figure=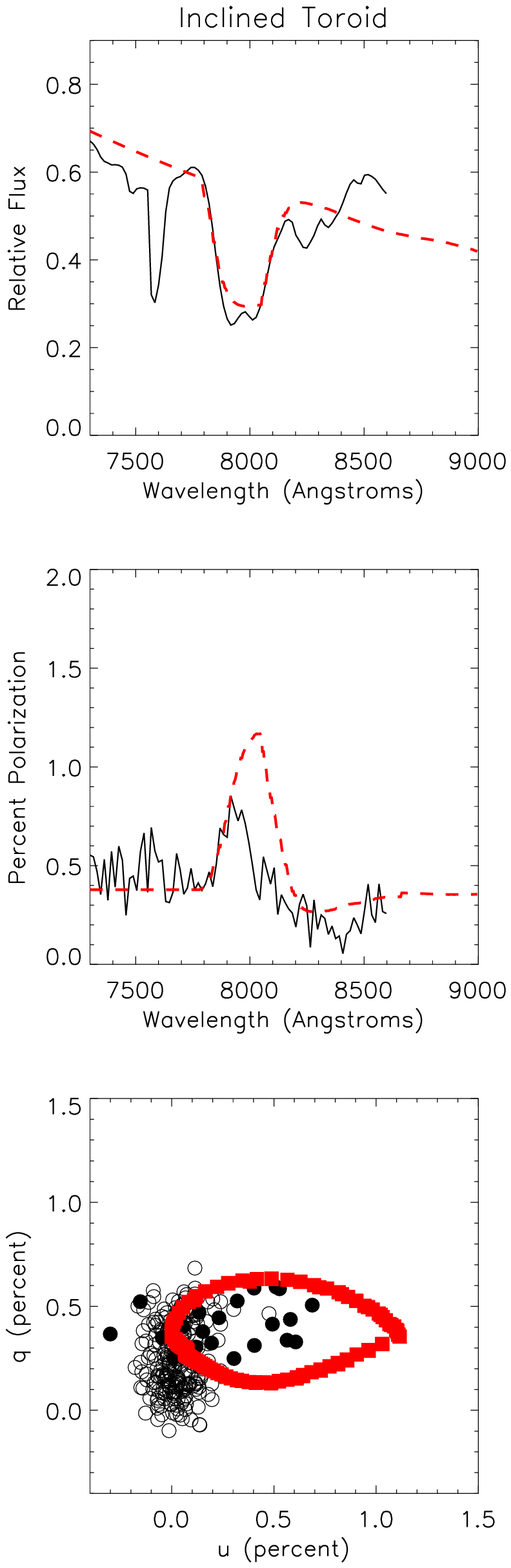,height=0.4\textheight}
      \end{center}
  \end{minipage}
  \hfill
\caption{
The left half shows velocity slices through an edge-on torus 
surrounding an ellipsoidal photosphere, both expanding homologously.
This geometry breaks the axisymmetry.  The torus has a strong 
absorption line that blocks the continuum emission from the 
photosphere. The geometry of the blocking region depends on the 
velocity slices through the structure and hence on the wavelength
across the line profile. The arrows mark the photospheric polarization 
angle and amplitude. The three panels show decreasing blue shift.
The right half shows the effect of the torus, now tilted both with 
respect to the line of sight and to the axis of the underlying 
ellipsoidal photosphere, on the total flux line feature, the 
polarization feature, and as depicted in the Q/U plane. 
The polarization is somewhat strong compared to the particular 
observations of the Type Ia supernova SN~2001el (\S \ref{hiVCa}),
but the locus in the Q/U plane shows a loop that matches the
observations rather well. From Kasen et al. (2003). 
}
\end{figure}

\clearpage

\begin{figure}
  \label{87A_Vpol}  
  \centerline{\psfig{figure=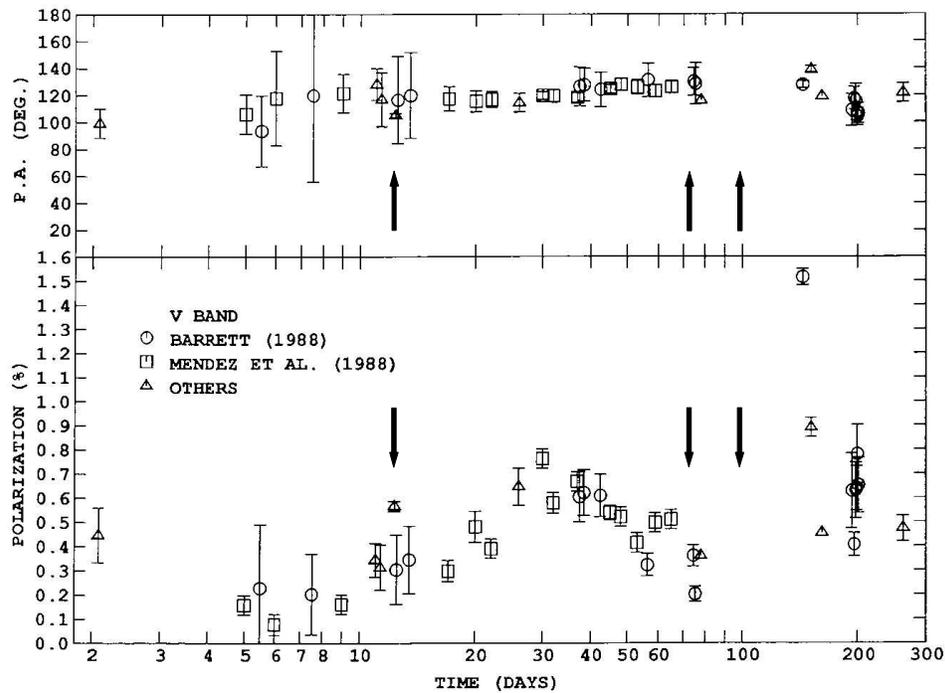,angle=-90,height=0.4\textheight}}
  \caption{Evolution of the V-band polarization
  and associated polarization angle of SN 1987A from Jeffery (1991a).
  Note the near constancy of the polarization angle despite the
  variation in the amplitude of the broad band polarization. Vertical
  arrows indicate the epochs for which spectropolarimetry is 
  presented in Figures 4 and 5.
  }
\end{figure}

\clearpage

\begin{figure}[h]
  \label{87A_0505}
  \hfill
\begin{minipage}[t]{.45\textwidth}
      \begin{center}
      \psfig{figure=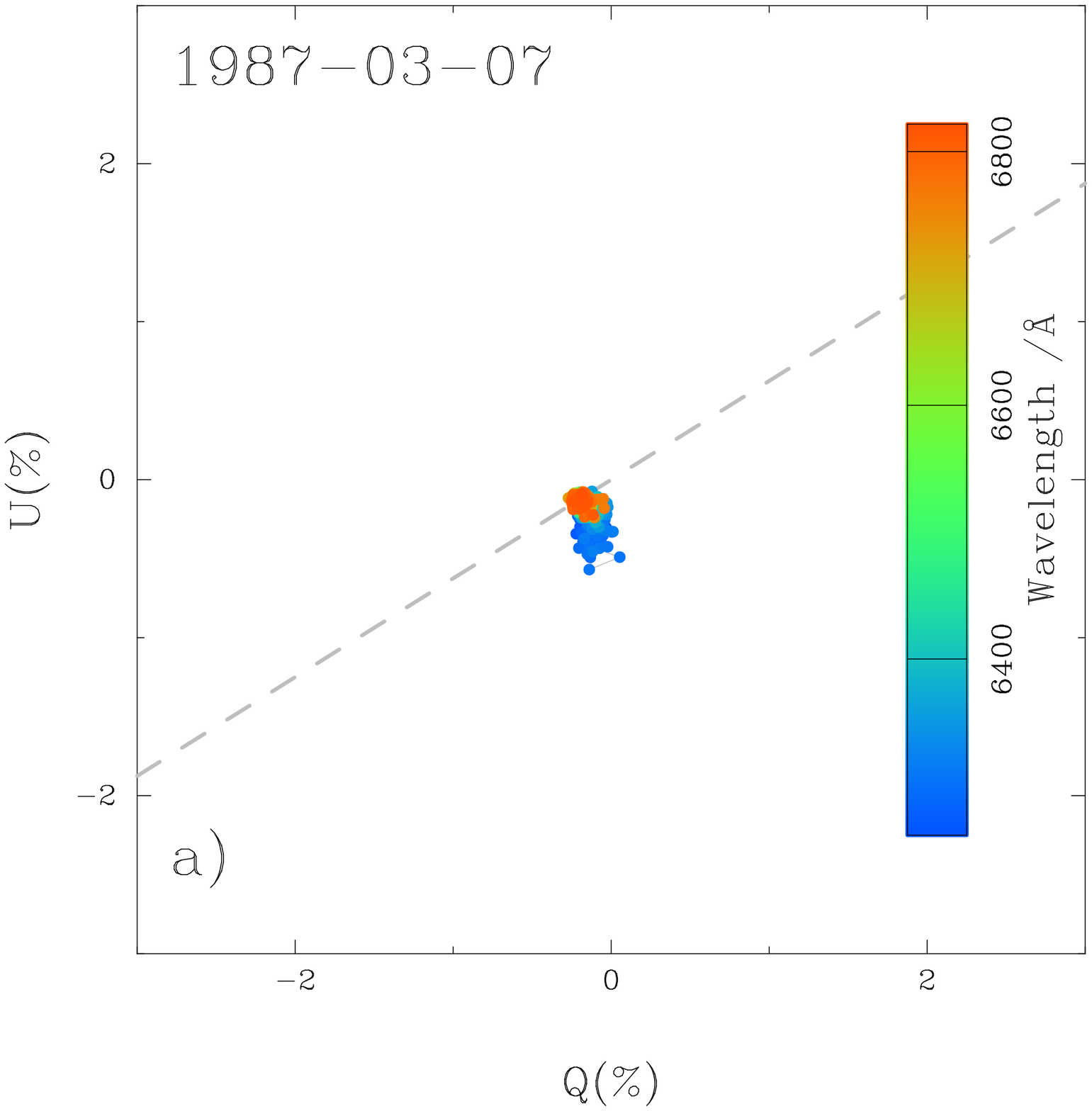,height=0.25\textheight,angle=-90}
      \end{center}
  \end{minipage}
  \hfill
  \begin{minipage}[t]{.45\textwidth}
      \begin{center}
      \psfig{figure=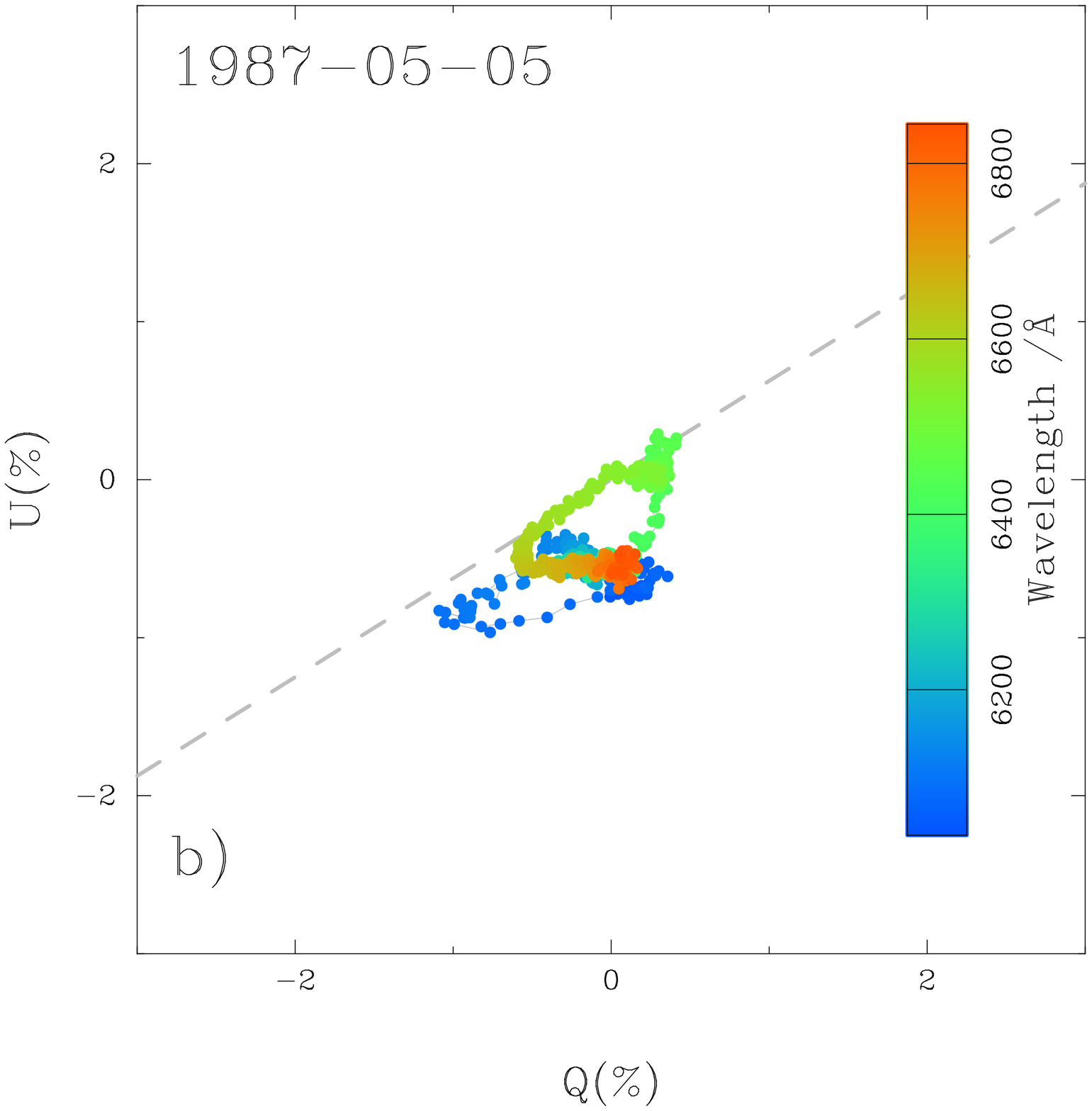,height=0.25\textheight,angle=-90}
      \end{center}
  \end{minipage}
  \hfill
\caption{(Left Panel) H$\alpha$ and adjacent continuum in the 
  Q/U plane for SN~1987A on March 7, 1987, about two weeks after 
  the explosion. The data cluster near the origin with some small 
  departure to negative U at a position angle of about 150\degree. 
  (Right Panel) H$\alpha$ and adjacent continuum on May 5, 1987, 
  near maximum light.  The H$\alpha$ polarization has developed a distinct 
  ``loop" structure indicating significant departures from axisymmetry 
  and bears little relation to the data of March 7. The solid line 
  corresponds to the ``speckle angle," $\theta \sim$ 16\degree\ (Meikle 
  et al. 1987) in the negative quadrant and extrapolated into the positive
  quadrant as if there were an oblate counterpart as a guide to the eye. 
  The data have been corrected for the ISP given by 
  M\'endez (1990; Jeffery 1991a).  Adapted from Cropper et al. (1988). 
  } 
\end{figure}

\clearpage
   
\begin{figure}[h]
  \label{87A_0603He}
  \hfill
\begin{minipage}[t]{.45\textwidth}
      \begin{center}
      \psfig{figure=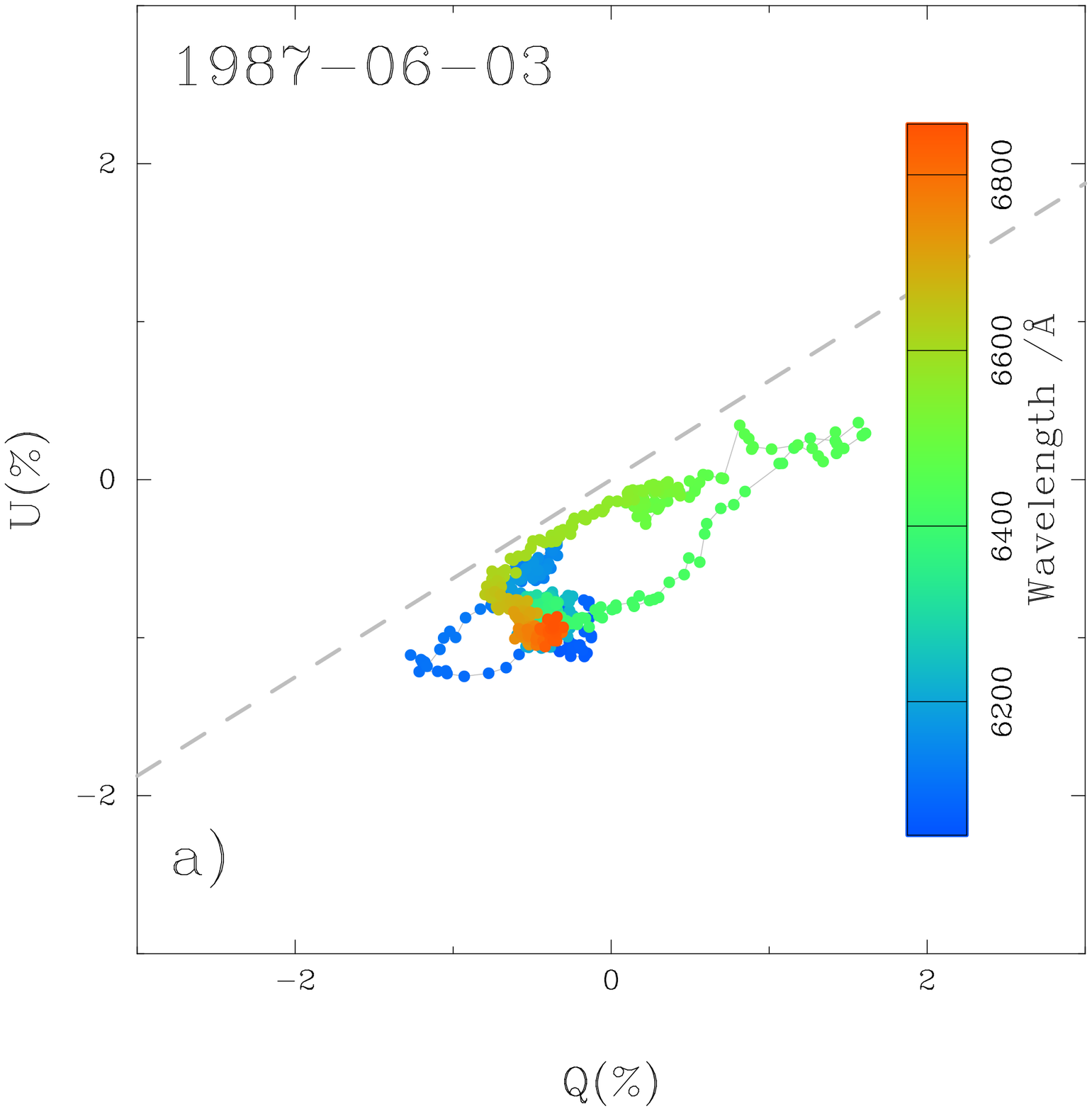,height=0.25\textheight,angle=-90}
      \end{center}
  \end{minipage}
  \hfill
  \begin{minipage}[t]{.45\textwidth}
      \begin{center}
      \psfig{figure=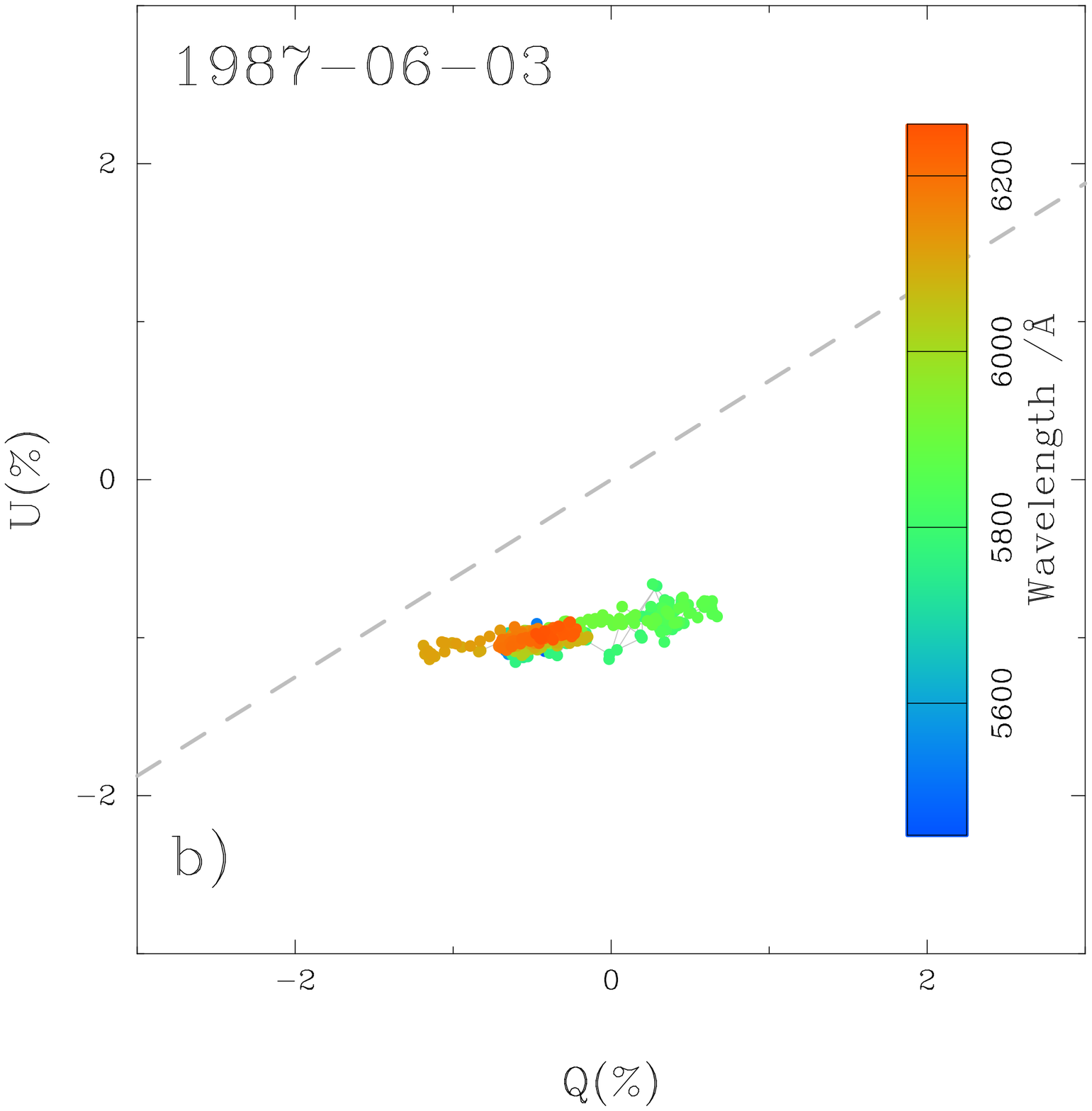,height=0.25\textheight,angle=-90}
      \end{center}
  \end{minipage}
  \hfill
\caption{(Left Panel) H$\alpha$ and adjacent continuum on the Q/U 
  plane for SN~1987A on June 28, 1987, as the fading from peak began. 
  The ``loop" structure is now especially prominent.  (Right Panel) 
  He I $\lambda$ 5876 and adjacent continuum on June 28, 1987. The 
  data at around 5600 and 6200 \AA\ represent the continuum. The He 
  I line shows a large excursion to Q $\sim$ 0.7\% at 5800 \AA\ at 
  absorption minimum. The data at Q $\sim$ -1.2\% at 6100 \AA\ 
  corresponds to the absorption minimum of the adjacent Ba II line.  
  All these features fall closely along the same locus.  The solid 
  line corresponds to the ``speckle angle," $\chi \sim$ 16\degree\ 
  (Meikle et al. 1987) in the negative quadrant and extrapolated into 
  the positive quadrant as if there were an oblate counterpart as a 
  guide to the eye. The data have been corrected for the ISP 
  given by M\'endez (1990; Jeffery 1991a).  Adapted from Cropper 
  et al. (1988).
}
\end{figure}
 
\clearpage
 
\begin{figure}
  \label{sn1999em}
  \centerline{\psfig{figure=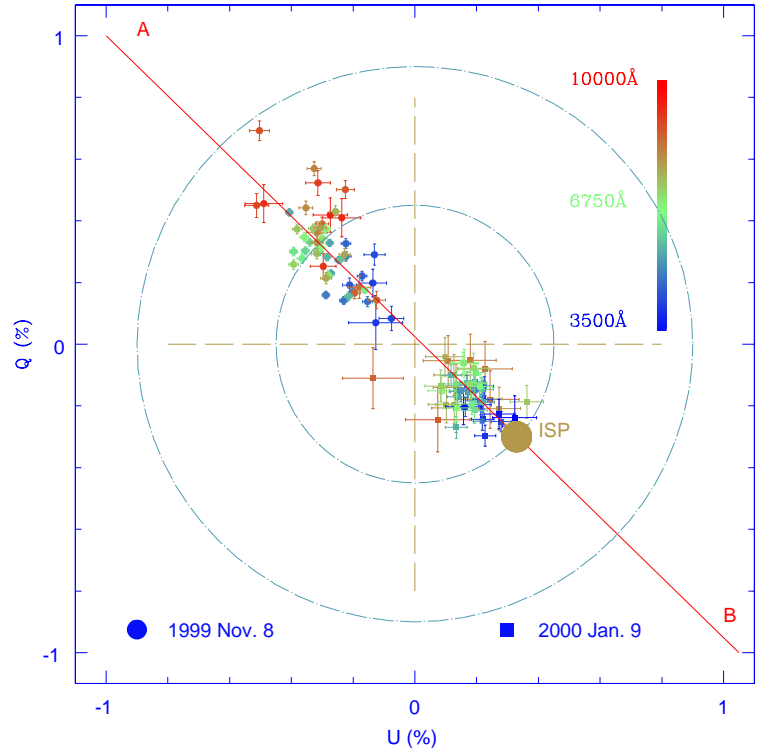,height=0.5\textheight}}
\caption{
Polarimetry of the Type II SN 1999em. The wavelengths represented 
by different data points are color encoded.  The 1999 2
Nov. data (clustered in the lower-right quadrant) corresponding to
the early plateau are clearly separated from the 2000 9 Jan. data
(clustered in the upper-left quadrant) on the late plateau,
showing an increase in polarization with time.  Line AB is the
dominant axis. The circles are the upper limits to the
interstellar polarization (measured with respect to the observed values Q = 0, U = 0) assuming E(B-V) toward the supernova
to be 0.05 (inner circle) and 0.1 (outer circle). The approximate
location of the component due to interstellar dust is shown as
a solid circle.
}
\end{figure}

\clearpage

\begin{figure}[h]
\label{sn2004dj}
  \hfill
\begin{minipage}[t]{.45\textwidth}
      \begin{center}
      \psfig{figure=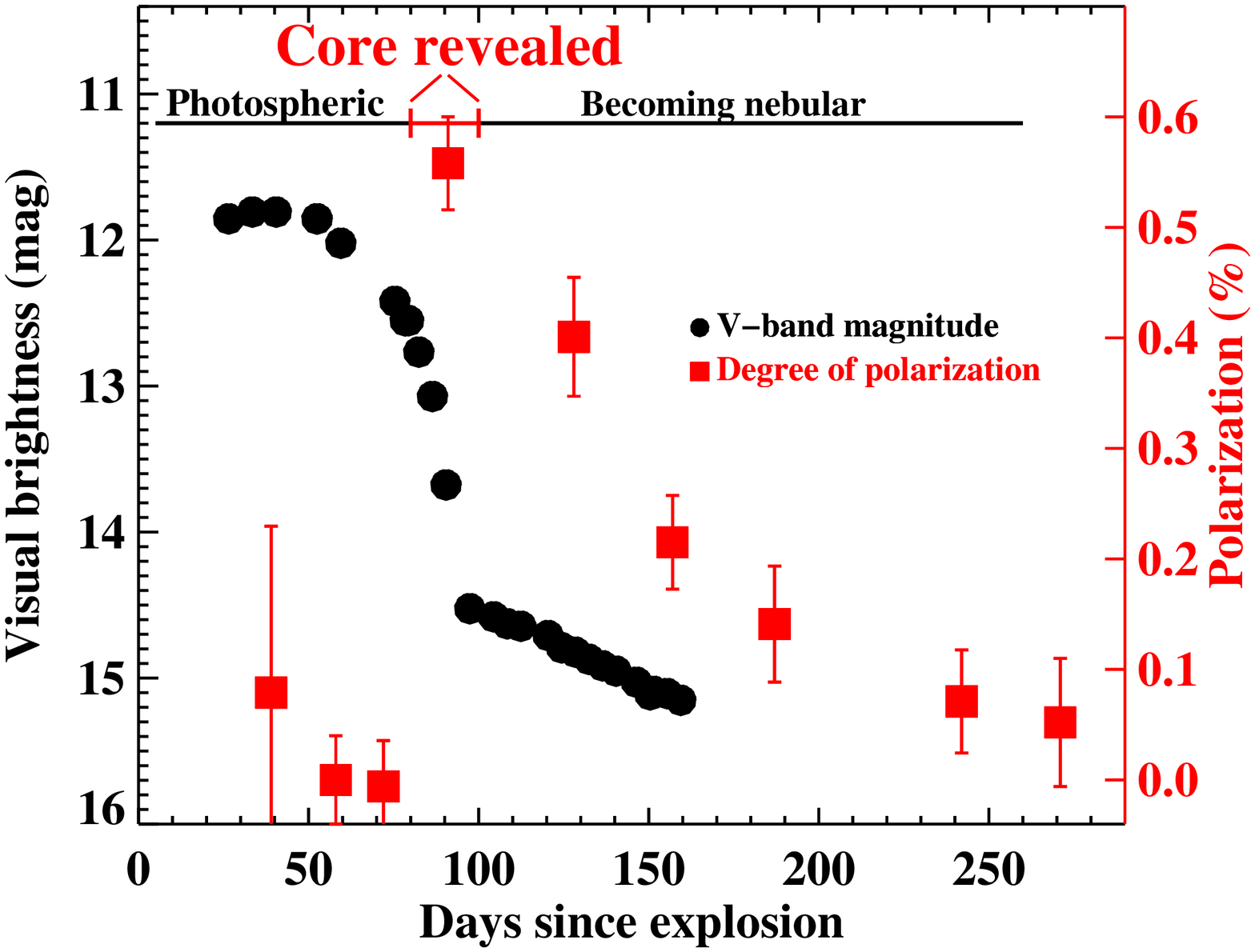,height=0.2\textheight,angle=90}
      \end{center}
  \end{minipage}
  \hfill
  \begin{minipage}[t]{.45\textwidth}
      \begin{center}
      \psfig{figure=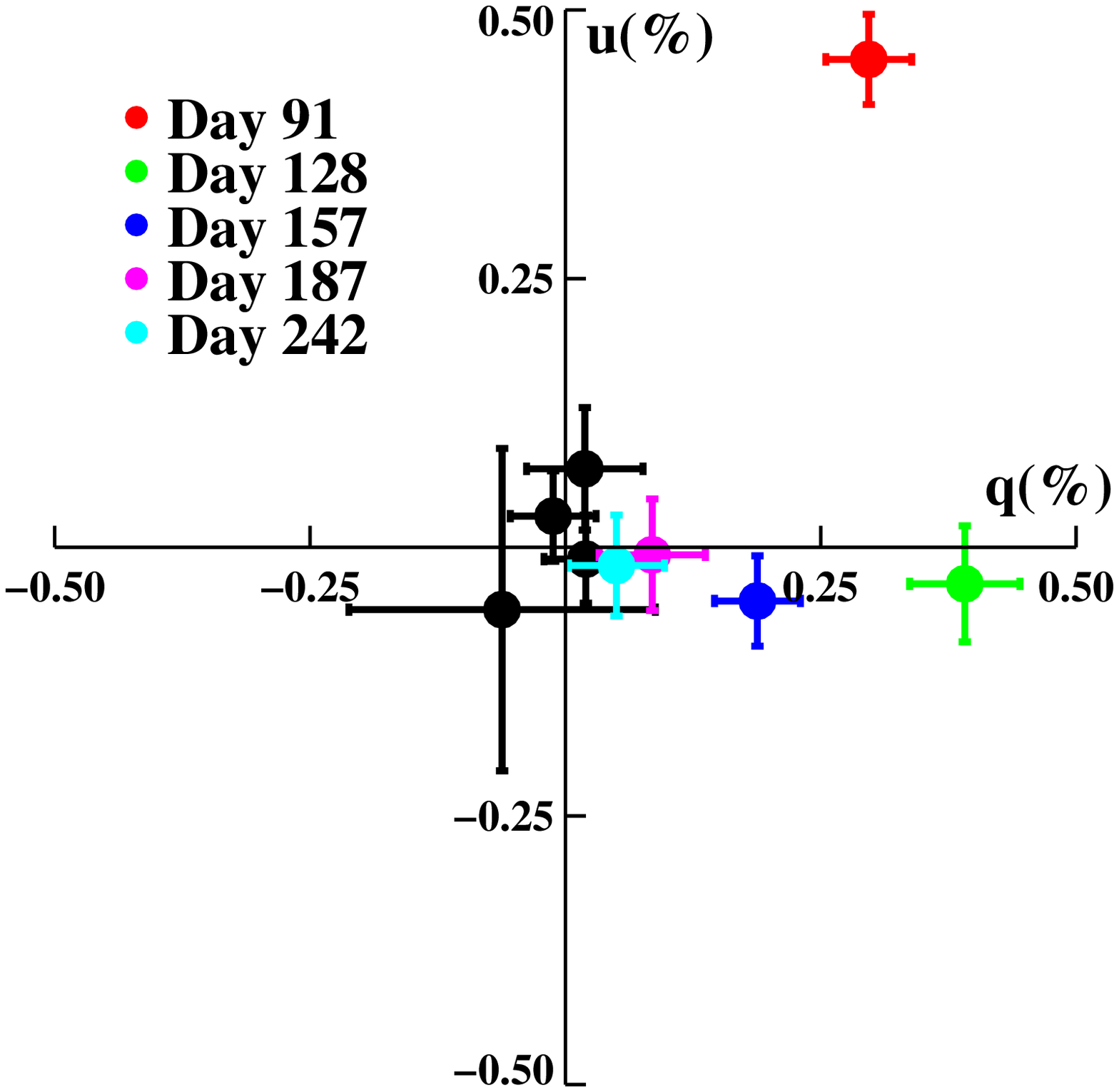,height=0.2\textheight,angle=0}
      \end{center}
  \end{minipage}
  \hfill
\caption{(Left Panel) Light curve and continuum 
polarization of SN 2004dj. 
The red points represent the median of the rotated Stokes parameter in the
ISP-subtracted data over the spectral range 6800 - 8200 \AA. From
Leonard et al. (2006). (Right Panel) Evolution of SN~2004dj with
time in the Q/U plane (courtesy D. Leonard).   
}
\end{figure}

\clearpage

\begin{figure}[h]
\label{sn2001ig_loops}
  \hfill
\begin{minipage}[t]{.45\textwidth}
      \begin{center}
      \psfig{figure=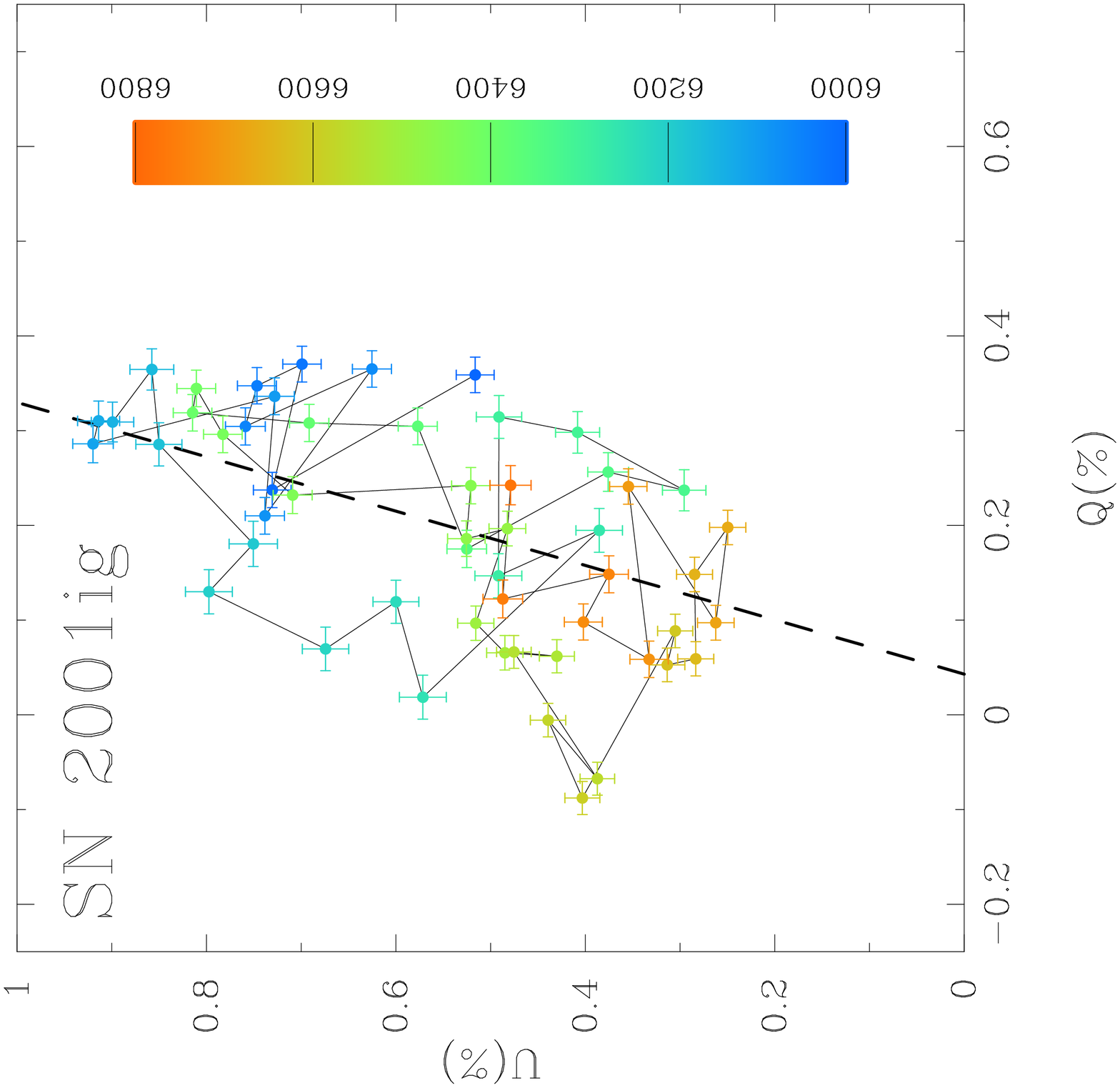,height=0.3\textheight,angle=-90}
      \end{center}
  \end{minipage}
  \hfill
  \begin{minipage}[t]{.45\textwidth}
      \begin{center}
      \psfig{figure=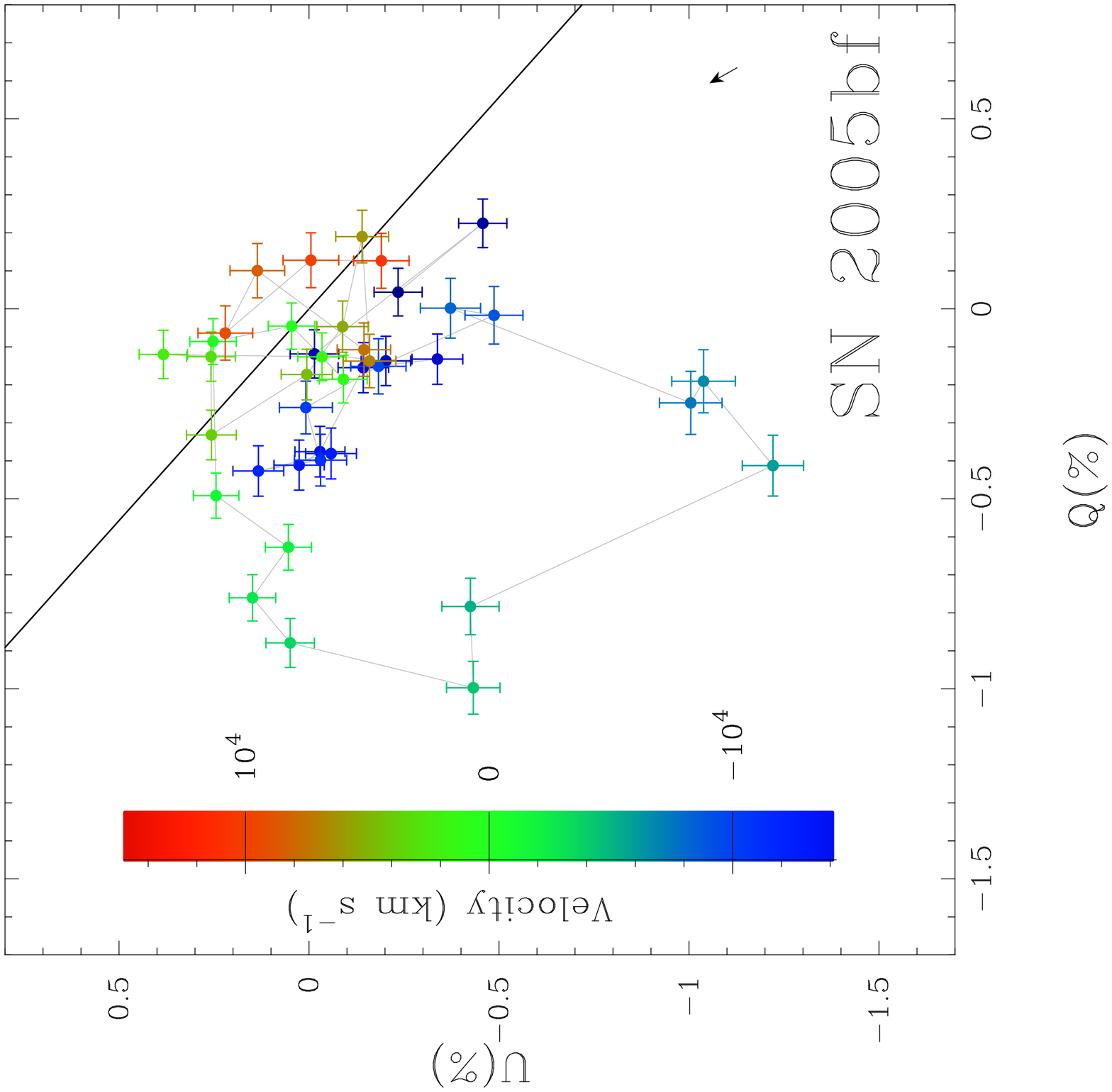,height=0.3\textheight,angle=-90}
      \end{center}
  \end{minipage}
  \hfill
  \caption{Q/U plane loops, corrected for the relevant
  ISP, of He I 6678\AA/H$\alpha$\ in the Type IIb SN 2001ig 
  (Left Panel: Maund et al. 2007a) and He I 5876\AA\ in the 
  Type Ib/c SN 2005bf (Right Panel: Maund et al. 2007b).
  The loops are indicative of non-axisymmetric structure.}
\end{figure}

\clearpage

\begin{figure}
  \centerline{\psfig{figure=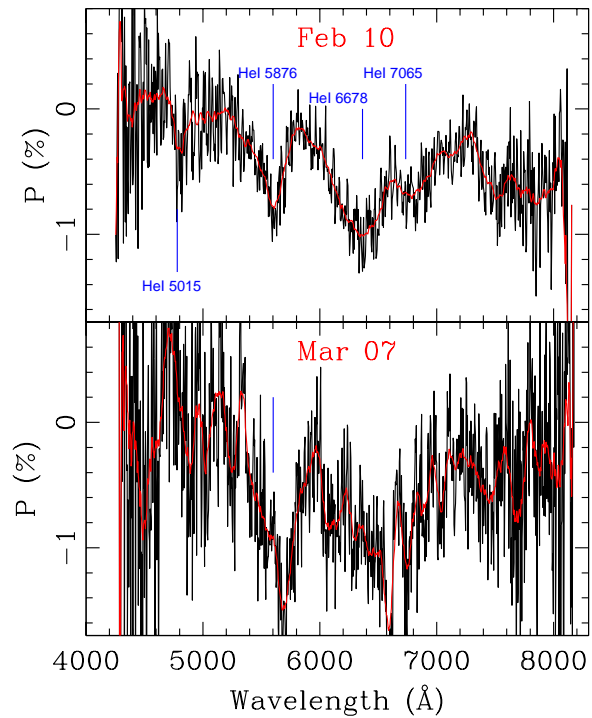,height=0.5\textheight}}
  \caption{Polarization spectrum of Type Ic SN~1997X on 10 February 
and 7 March, 1997 approximately 5 and 30 days after maximum light. 
Note the prominant evidence for high velocity He I lines in the 
10 February spectrum even though this event that was otherwise 
identified as being helium deficient.
}
\end{figure}

\clearpage

\begin{figure}
\label{sn2001el}
  \centerline{\psfig{figure=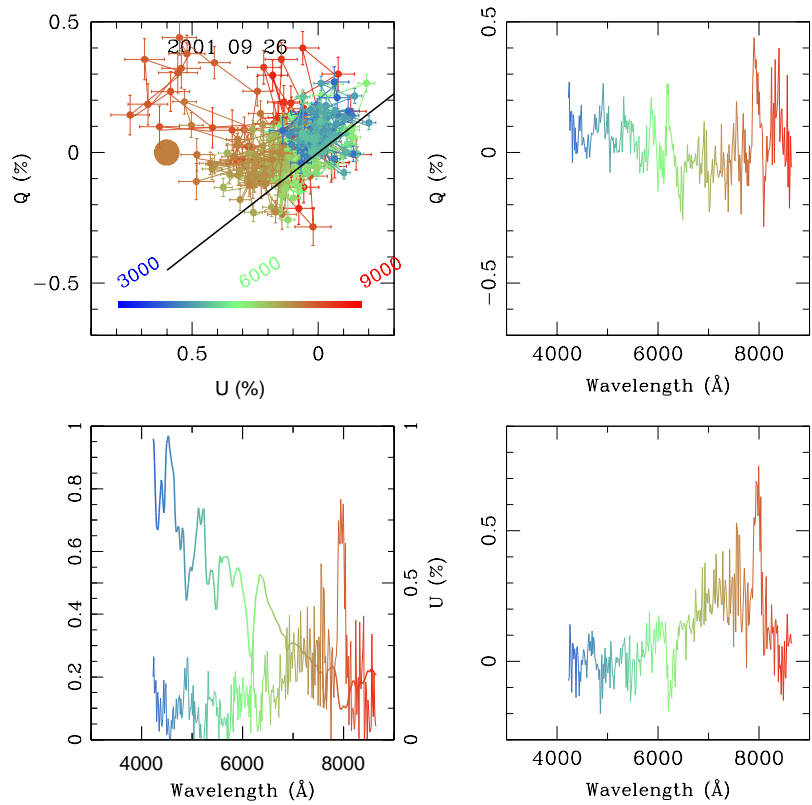,height=0.5\textheight}}
  \caption{Spectropolarimetry of the Type Ia SN 2001el on 2001 Sept. 26, 
7 days before maximum. The assumed interstellar polarization is shown 
as a solid dot in the Q/U plot (upper left).  The straight line 
illustrates the dominant axis shifted to the origin of the Q/U plot.  
The Q (upper right) and U (lower right) spectra show conspicuously 
polarized  spectral features. The flux spectrum is also given in
the lower left panel as the line rising to the left. The lower
left panel shows the correlations of the degree of polarization and 
the spectral features. The high-velocity Ca II IR feature is 
especially prominant at 8000 \AA. The wavelength color code is 
presented at the bottom of the upper left panel.  From Wang et al. (2003a).
}
\end{figure}

\clearpage

\begin{figure}
\label{sn2004dtSiO}
  \centerline{\psfig{figure=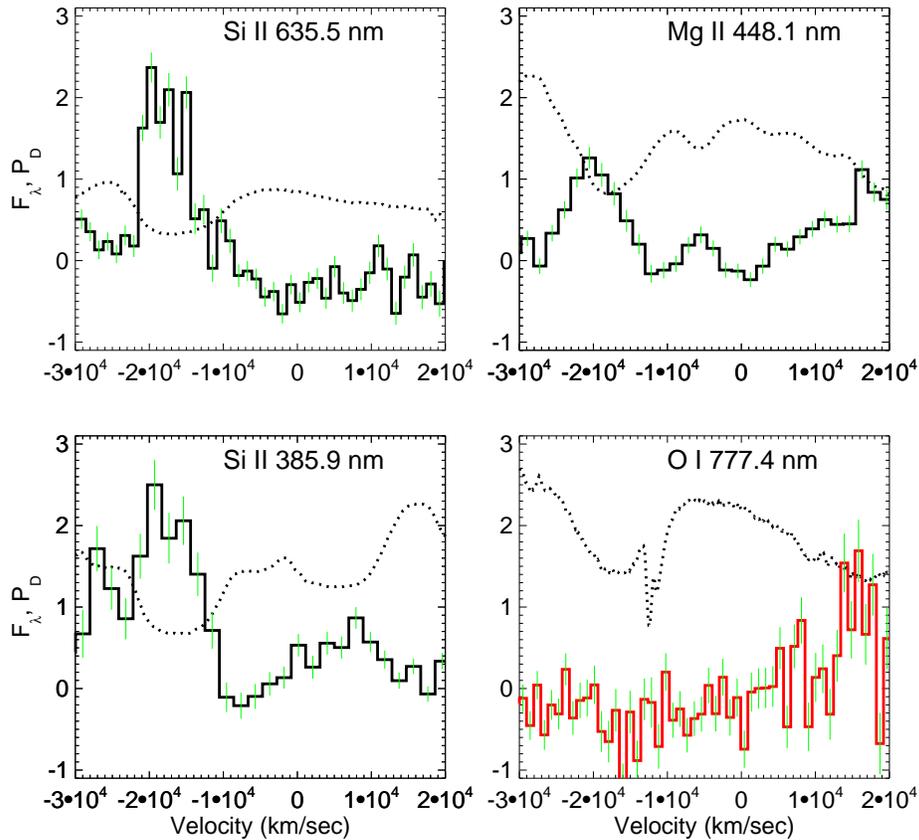,height=0.5\textheight}}
  \caption{
Polarized line features in SN~2004dt.
Absorption features in the total flux spectrum and the polarization
profiles along the dominant axis are given as a function of
velocity for the lines of (clockwise from the upper left) Si II 
$\lambda$6355, Mg II $\lambda$4471, O I $\lambda$7774, and 
Si II $\lambda$3859. 
The dotted lines give the total flux profiles and the solid
lines give the polarization. Note that there is significant
polarization in all the lines except O I. From Wang et al. (2006).  
} 
\end{figure}

\begin{figure}
\label{SN04dtQU}
  \centerline{\psfig{figure=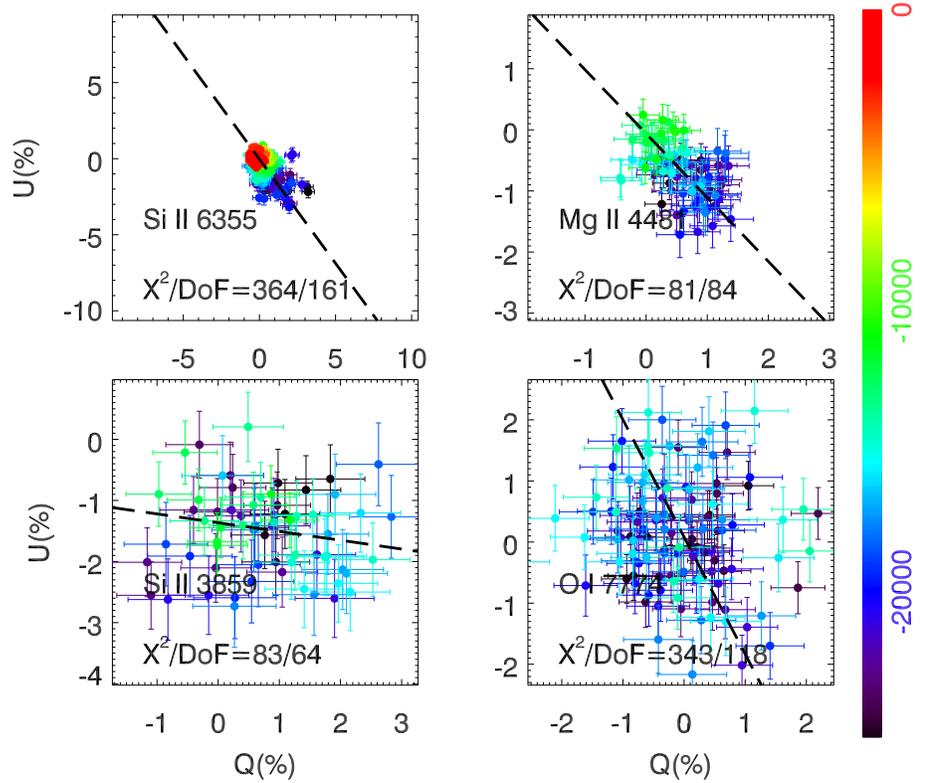,height=0.5\textheight}}
  \caption{ Q-U diagrams of the strongest lines in SN~2004dt and the
corresponding linear fits given by the dashed lines. The ranges of the
data points  are from -25,000 to -10,000 \kms\ for these lines. The
panels are for the lines of Si II $\lambda$6355 (top left), 
Mg II $\lambda$4481 (top right), O I $\lambda$7774 (bottom right), 
and Si II $\lambda$3859 (bottom left). Only the Mg II
line is consistent with a straight line, indicating a simple
axially symmetric geometry with no detectable clumping.
From Wang et al. (2006).
} 
\end{figure}

\begin{figure}
\label{SiII6355}
  \centerline{\psfig{figure=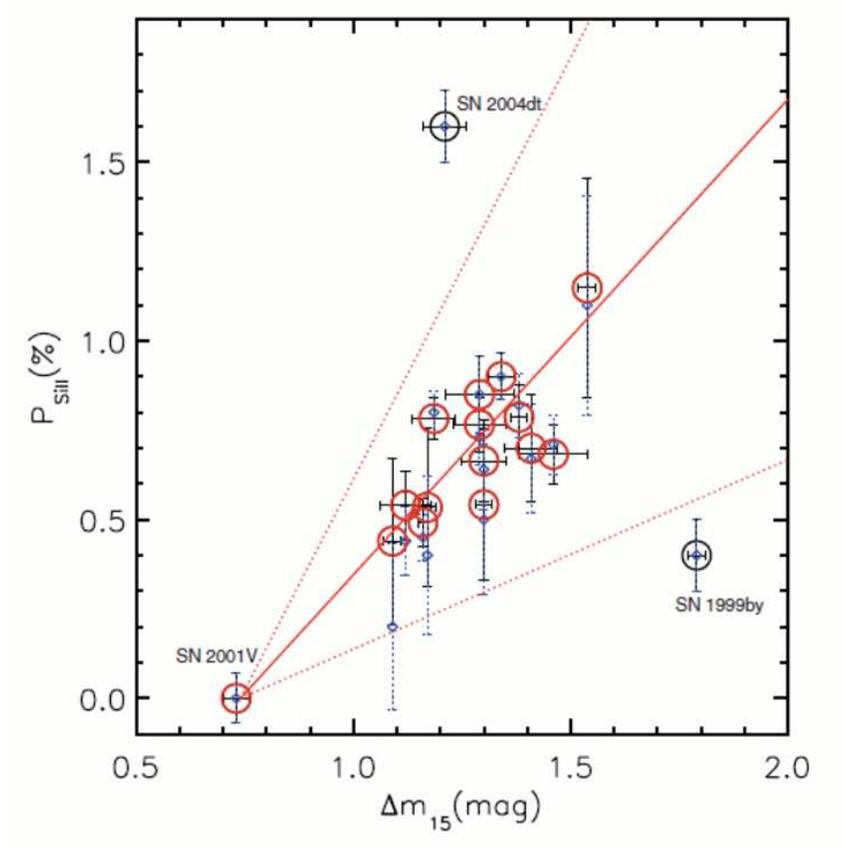,height=0.5\textheight}}
  \caption{
 The correlation between the degree of polarization across
the Si II $\lambda$6355 line and the light curve decline rate,
$\Delta\rm{m_{15}}$, for a
sample of 17 Type Ia supernovae corrected to 5 days after maximum.
The linear fit represented by the straight line includes only
spectroscopically normal supernovae shown as red open circles.
The blue open circles show the highly polarized event SN~2004dt
and the subluminous SN~1999by. The dotted lines illustrate the 1
$\sigma$ level of the intrinsic polarization distribution around
the most likely value for the Monte Carlo simulation with 20
identical opaque clumps of Si II. From Wang, Baade, \& Patat (2007).
} 
\end{figure}

\end{document}